\documentclass[sigconf,10pt]{acmart}
\renewcommand\footnotetextcopyrightpermission[1]{}
\usepackage{enumitem}
\usepackage{microtype} 
\usepackage{xspace}
\usepackage{pifont}
\usepackage{booktabs}
\usepackage{tabularx}
\usepackage{multirow}
\usepackage[linesnumbered, ruled]{algorithm2e}
\usepackage{algpseudocode}
\usepackage{graphicx}
\usepackage{subfigure}
\usepackage{caption} 
\usepackage[most]{tcolorbox}
\usepackage{wasysym}
\usepackage{fp}
\usepackage{colortbl}
\newlength\maxlenbar
\newcommand\tablebar[3][blue!15]{%
  \FPeval\result{round((#3)/#2:4)}%
  \rlap{\textcolor{#1}{\hspace*{\dimexpr-\tabcolsep+.5\arrayrulewidth}%
        \rule[-.05\ht\strutbox]{\result\maxlenbar}{.95\ht\strutbox}}}%
  \makebox[\dimexpr\maxlenbar-0.2\tabcolsep+\arrayrulewidth][r]{#3}}
\newcommand\deltabar[3][red!15]{%
  \FPeval\result{round((#3)/#2:4)}%
  \rlap{\textcolor{#1}{\hspace*{\dimexpr-\tabcolsep+.5\arrayrulewidth}%
        \rule[-.05\ht\strutbox]{\result\maxlenbar}{.95\ht\strutbox}}}%
  \makebox[\dimexpr\maxlenbar-0.2\tabcolsep+\arrayrulewidth][r]{$-$#3}}
\def\headerbar{Macro-F1 (\%)}
\newcommand{\SystemName}{\texttt{RAGent}\xspace}

\begin{document}

\title{RAGent: Physics-Aware Agentic Reasoning for Training-Free mmWave Human Activity Recognition}

\author{Mingda Han$^{1}$, Huanqi Yang$^{2}$, Zehua Sun$^{3}$, Wenhao Li$^{2}$, Yanni Yang$^{1}$,\\ Guoming Zhang$^{1}$, Yetong Cao$^{1}$, Weitao Xu$^{2}$, Pengfei Hu$^{1}$}
\affiliation{
\institution{$^1$Shandong University, $^2$City University of Hong Kong, $^3$National University of Singapore
}
\country{}
}
\renewcommand{\shortauthors}{Han et al.}

\begin{abstract}
Millimeter-wave (mmWave) radar enables privacy-preserving human activity recognition (HAR), yet real-world deployment remains hindered by costly annotation and poor transferability under domain shift. 
Although prior efforts partially alleviate these challenges, most still require retraining or adaptation for each new deployment setting. 
This keeps mmWave HAR in a repeated collect--tune--redeploy cycle, making scalable real-world deployment difficult.
In this paper, we present \SystemName, a deployment-time training-free framework for mmWave HAR that reformulates recognition as evidence-grounded inference over reusable radar knowledge rather than deployment-specific model optimization.
Offline, \SystemName constructs a reusable radar knowledge base through constrained cross-modal supervision, where a Vision-Language Model (VLM) transfers activity semantics from synchronized videos to paired radar segments without manual radar annotation. 
At deployment time, \SystemName recognizes activities from radar alone by retrieving physically comparable precedents in an explicit kinematic space and resolving the final label through structured multi-role reasoning.
The reasoning protocol is further refined offline through zero-gradient self-evolution.
Extensive experiments on a self-collected dataset show that \SystemName achieves 93.39\% accuracy without per-domain retraining or target-domain adaptation, while generalizing robustly across domains.
\end{abstract}

\maketitle

\settopmatter{printfolios=true}
\pagestyle{plain} 

\vspace{-0.2in}
\section{Introduction}
\subsection{Background and Motivation}

\begin{figure}[t]
\centering
\includegraphics[width=\columnwidth]{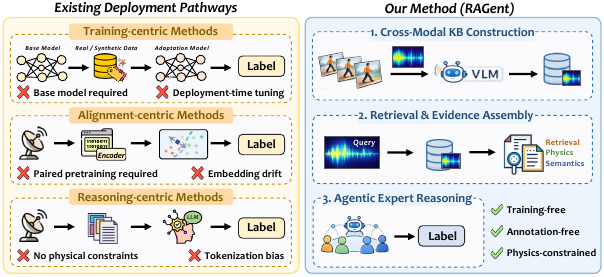}
\vspace{-0.2in}
\caption{\textbf{Paradigm shift in mmWave HAR.} \SystemName moves beyond conventional methods by grounding recognition in retrieved precedents and verifiable kinematic evidence through agentic reasoning.}
\label{fig:motivation}
\vspace{-0.25in}
\end{figure}

Human activity recognition (HAR)~\cite{gu2021survey} is a key enabler for smart healthcare, assisted living, home rehabilitation, and ambient intelligence, supporting applications such as fall detection~\cite{alam2022vision}, recovery assessment~\cite{liao2020review}, and daily behavior understanding~\cite{debes2016monitoring}.
While vision-based solutions~\cite{beddiar2020vision} have achieved strong performance, they are often undesirable in privacy-sensitive spaces, and their reliability can be fragile under adverse lighting.
Millimeter-wave (mmWave) radar offers a compelling alternative: it is contactless, robust to lighting and common indoor conditions, and more privacy-preserving than cameras~\cite{zhang2023survey,gong2025seradar}.
Commercial products already demonstrate these capabilities, such as the Caaresys in-cabin system~\cite{caaresys} for contactless respiration and heart-rate monitoring and the Aqara Presence Sensor FP300 for human presence detection in smart homes~\cite{aqara}. 

A central challenge in mmWave HAR is how to represent human motion in a form that is both discriminative and reusable.
Prior work has mainly explored two representation families: \emph{spatial-structure-centric} representations (e.g., radar point clouds~\cite{hao2025mm,ding2024milliflow}) and \emph{velocity-signature-centric} representations (e.g., radar spectrograms~\cite{kim2022radar,feng2025mmwave}).
Point clouds are visually interpretable but are often limited by the angular resolution of commodity radars~\cite{zheng2024enhancing,zhao2020m}, making subtle motions hard to recover.
Radar spectrograms, in contrast, provide dense time-frequency velocity signatures and are particularly effective for fine-grained activity.
However, this strength comes with two long-standing bottlenecks that have collectively hindered real-world deployment.

\textbf{Semantic gap and the annotation bottleneck.}
Unlike RGB videos, radar spectrograms do not expose human-interpretable spatial semantics. Instead, they appear as energy patterns over time and frequency, which are information-rich yet highly non-intuitive. 
As a result, reliable radar annotation often requires expert effort, making large-scale dataset construction difficult and expensive.

\textbf{Domain shift and the retraining burden.}
Supervised mmWave HAR models are notoriously brittle under domain shift~\cite{zhang2023survey,kong2024survey}.
Even modest shifts can alter the signal distribution and cause substantial performance degradation. 
In practice, moving to a new domain often triggers a costly loop of
\emph{re-collection $\rightarrow$ re-labeling $\rightarrow$ retraining $\rightarrow$ re-validation}, which severely limits scalability.

A natural question is therefore whether mmWave HAR can move beyond the repeated \emph{collect--label--train} cycle and become deployable without per-domain optimization. 
Existing efforts provide partial answers, but they largely follow three deployment pathways, as summarized in Fig.~\ref{fig:motivation} (left). 
\emph{Training-centric methods} improve transferability through adaptation~\cite{liu2022mtranssee,wang2023rf,ding2020rf} or synthetic/generative data~\cite{zhang2022synthesized,chen2023rfgenesis,hu2025human}, yet still rely on training a better recognizer for the target deployment domain. 
\emph{Alignment-centric methods}~\cite{cao2024mmclip,yan2025mmexpert} avoid explicit supervised classification by aligning radar observations with semantic spaces, but their decisions remain tied to learned embedding similarity that can drift across domains. 
\emph{Reasoning-centric methods}~\cite{lai2025radarllm} introduce flexible Vision-Language Model (VLM)/Large Language Model (LLM)-based interpretation of radar observations, but without explicit physical grounding, they can still produce fluent yet physically inconsistent predictions. 
What remains missing is therefore not simply a more label-efficient or more powerful recognizer, but a different deployment-time paradigm: one that avoids manual radar annotation and per-domain optimization, while grounding decisions in physically verifiable evidence rather than domain-sensitive latent similarity.

In this paper, we present \SystemName, a deployment-time training-free framework for mmWave HAR that fundamentally shifts the paradigm from deployment-specific model optimization to evidence-grounded inference via reusable radar knowledge, as illustrated in Fig.~\ref{fig:motivation} (right). 
Instead of adapting a classifier to each new domain, \SystemName acquires reusable radar knowledge once from synchronized video--radar pairs without manual radar labeling, and then reuses it at deployment time to resolve new queries through physically comparable precedents and constrained multi-role reasoning under physical consistency. 
Through this shift, the basis of generalization moves from learning a better deployment-specific recognizer to reusing historical evidence, explicit kinematic descriptors, and structured arbitration.

\subsection{Challenges and Contributions}

\begin{figure}[t]
    \centering
    \subfigure[Semantic asymmetry.]{
        \includegraphics[width=0.31\linewidth]{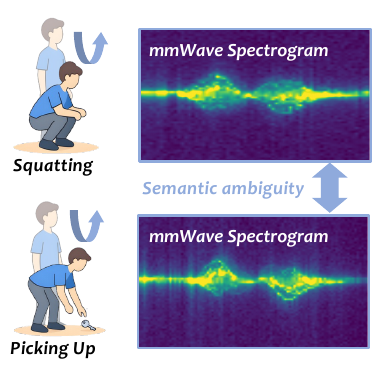}
        \label{fig:challenge1}
    }
    \subfigure[Unstable reasoning under domain shift.]{
        \includegraphics[width=0.62\linewidth]{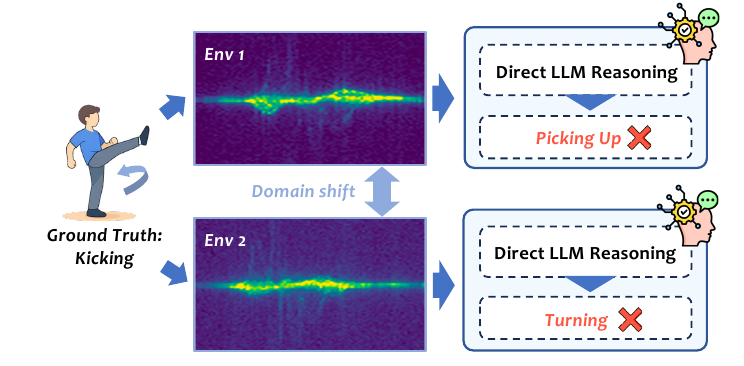}
        \label{fig:challenge2}
    }
    \vspace{-0.15in}
    \caption{\textbf{Two key challenges in training-free mmWave HAR.}}
    \label{fig:challenge}
    \vspace{-0.4in}
\end{figure}

Realizing \SystemName requires addressing two key challenges, as illustrated in Fig.~\ref{fig:challenge}.

\textbf{Challenge 1: Reliable semantic transfer for non-intuitive radar data.}
Distinct activities can exhibit similar mmWave spectrogram patterns, while the radar observations themselves remain difficult for humans to interpret directly. 
This semantic asymmetry makes large-scale manual radar annotation expensive and error-prone.
Although recent efforts have begun to connect mmWave observations with semantic information through cross-modal alignment or radar-language understanding~\cite{cao2024mmclip,yan2025mmexpert,lai2025radarllm}, they do not directly address how to build reliable knowledge-base entries without manual radar labeling.
A natural alternative is to transfer labels from synchronized videos using a VLM, but naive pseudo-labeling can introduce noisy supervision into the knowledge base. 
To address this challenge, \SystemName performs constrained cross-modal supervision that transfers semantics from synchronized video to paired radar segments while filtering unreliable pseudo-labels before they enter the reusable knowledge base.

\textbf{Challenge 2: Grounded training-free inference under domain shift.}
Even for the same activity class, radar signatures can vary noticeably across domains. 
Conventional approaches typically handle such variation through target-domain adaptation or retraining~\cite{liu2022mtranssee,wang2023rf}, but such optimization is not assumed in our deployment-time training-free setting.
The challenge is therefore how to make stable decisions from radar alone under domain shift. 
A natural alternative is to leverage foundation models to interpret radar signatures directly.
However, direct semantic interpretation of raw radar signatures can be unreliable across domains, and unconstrained language reasoning may further produce plausible but physically invalid predictions. 
Deployment-time training-free HAR therefore requires evidence that remains comparable across domains and verifiable under physical constraints. 
To address this challenge, \SystemName combines physics-driven precedent retrieval with structured multi-role reasoning, so that semantic observations are always resolved under retrieval support and physical consistency checks.

\begin{figure*}[t]
\centering
\includegraphics[width=0.99\linewidth]{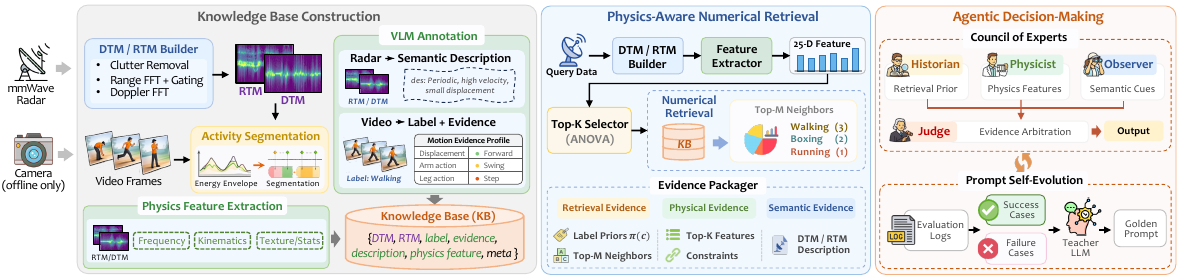}
\vspace{-0.15in}
\caption{\textbf{\SystemName overview.} The framework follows an offline-to-online workflow with three stages: (1) \emph{Knowledge-Base Construction}, 
(2) \emph{Physics-Driven Numerical Retrieval}, and 
(3) \emph{Council-of-Experts Reasoning}.}
\vspace{-0.1in}
\label{fig:overview}
\end{figure*}

These designs enable training-free mmWave HAR without restarting the full \emph{collect--label--train} loop in each new domain. This paper makes the following contributions:
\begin{itemize}[leftmargin=*]
\vspace{-0.05in}
    \item We present \SystemName, a deployment-time training-free framework for mmWave HAR that shifts the paradigm from deployment-specific model optimization to evidence-grounded inference over reusable radar knowledge and multi-role agentic reasoning.

    \item We design a unified offline-to-online pipeline that addresses both annotation and generalization bottlenecks: it constructs a reusable radar knowledge base through constrained cross-modal supervision without manual radar annotation, and performs deployment-time recognition through reusable precedents, explicit kinematic evidence, and physics-aware agentic reasoning.

    \item We evaluate \SystemName on a self-collected dataset of 2,568 mmWave segments from 8 participants, 3 environments, and 12 activities. \SystemName achieves 93.39\% accuracy without per-domain retraining or target-domain adaptation, remains robust under cross-environment, cross-subject, and cross-date settings, and generalizes across different LLM backends.
\end{itemize}

\section{Formulation and Principles}
\label{sec:principles}

\subsection{Problem Formulation}
\label{subsec:problem_formulation}

We formulate deployment-time mmWave HAR as a \emph{knowledge-grounded inference} problem rather than a deployment-specific model adaptation problem. 
In our setting, \emph{deployment-time training-free} means that no per-deployment radar recollection, manual re-labeling, task-specific retraining, fine-tuning, gradient-based adaptation, or human-in-the-loop radar annotation is performed for the target deployment domain. 
Instead, the system accesses a reusable knowledge base $\mathcal{B}$ constructed offline, where each entry stores a historical radar segment together with the associated physical and semantic evidence needed for future inference, while any protocol refinement is performed offline as a one-time pre-deployment procedure rather than target-domain adaptation.

Given a query radar segment $q$, the prediction is resolved from three evidence sources: the retrieved precedent set $\mathcal{N}_M(q)$ from $\mathcal{B}$, the explicit physical descriptors $\mathbf{p}(q)$ extracted from $q$, and the semantic observations $\mathbf{t}(\boldsymbol{M}_q)$ parsed from its radar feature maps $\boldsymbol{M}_q$. The final prediction can be written abstractly as
\begin{equation}
\label{eq:problem_formulation}
\footnotesize
\setlength\abovedisplayskip{0.15cm}
\setlength\belowdisplayskip{0.15cm}
\hat{y}=
\arg\max_{c\in\mathcal{C}}
\mathrm{Score}\!\left(
c \mid \mathcal{N}_M(q),\ \mathbf{p}(q),\ \mathbf{t}(\boldsymbol{M}_q)
\right),
\end{equation}
where the decision is made in an explicit evidence space rather than by deployment-time optimization, and $\mathrm{Score}(\cdot)$ abstracts the council-based label resolution process.

A practical solution under this formulation should satisfy two requirements: the knowledge base must be constructible without repeated manual radar labeling, and the final decision process must remain stable across domains while being grounded in physically verifiable evidence. These requirements motivate the following design principles.

\subsection{Design Principles}
\label{subsec:design_principles}

\textbf{Principle I: Semantic knowledge should be acquired once and reused across deployments.}
Because radar spectrograms are not semantically intuitive to human annotators, repeatedly recollecting and manually labeling radar data for each new deployment is impractical. A training-free system should therefore acquire semantic knowledge once through more interpretable supervision and transfer it to paired radar observations during offline construction. The goal is not merely to assign labels, but to build a reusable knowledge base that supports future recognition without deployment-time annotation.

\textbf{Principle II: Final decisions should be grounded in comparable and physically verifiable evidence.}
Eliminating deployment-time retraining alone is not sufficient if recognition still depends on domain-sensitive similarity or unconstrained semantic inference. To remain stable across domains, deployment-time decisions should be based on evidence that is both comparable across environments and quantitatively verifiable under physical constraints. In this way, new queries can be resolved against historical precedents in a stable and auditable manner.

These two principles motivate the three core components of \SystemName: offline knowledge-base construction, physics-aware retrieval, and structured multi-expert reasoning.
\vspace{-0.1in}

\section{System Design}

\subsection{Overview}
Fig.~\ref{fig:overview} shows the offline-to-online workflow of \SystemName, which consists of three stages: (1) offline knowledge-base construction from synchronized video-radar pairs without manual radar labeling, (2) online retrieval of physically similar precedents in an explicit kinematic feature space, and (3) final label resolution by a Council of Experts that integrates retrieval priors, semantic observations, and physical constraints.

\vspace{-0.1in}
\subsection{Knowledge-Base Construction}
\label{sec:kb}

\SystemName first builds a reusable radar knowledge base from synchronized video--radar pairs for subsequent retrieval and reasoning. 
Each candidate segment is first assigned an acceptance status during knowledge-base curation. Each accepted knowledge-base entry then stores its radar observations (DTM and RTM), a physics-driven feature vector, transferred pseudo-labels, and semantic metadata. 
The knowledge base thus serves as a reusable evidence store for both precedent retrieval and downstream multi-role reasoning.

\subsubsection{Radar Representation Construction}
\label{subsubsec:RepConstruction}

We first convert the raw mmWave data into two complementary representations: a Doppler-Time Map (DTM) and a Range-Time Map (RTM). The received baseband samples are organized into a complex radar cube $\boldsymbol{S}\in\mathbb{C}^{N_F\times N_I\times N_S}$, where $N_F$, $N_I$, and $N_S$ denote the numbers of frames, chirps per frame (slow time), and ADC samples per chirp (fast time), respectively. To suppress stationary clutter, we subtract the mean across chirps within each frame and then perform a range FFT along the fast-time dimension. Next, we apply subject-centric range gating to select a range of interest (ROI): a dominant subject peak is identified for each chirp and stabilized over time to obtain a robust center bin $p_c$. We define the ROI $\boldsymbol{H}$ as an $L$-bin window around $p_c$ and mask out all other bins.

Next, we perform a Doppler FFT along the slow-time dimension within the selected range region $\boldsymbol{H}$ to obtain a frame-level Range-Doppler map (RDM). Let $\boldsymbol{M}(n,i,j)$ denote the magnitude at frame $n$, Doppler bin $i\in\{1,\dots,N_D\}$, and range bin $j\in\{1,\dots,N_R\}$. Based on $\boldsymbol{M}$, we construct two complementary 2D representations. The DTM summarizes velocity evolution by aggregating Doppler responses over the range axis:
\begin{equation}
\footnotesize
\setlength\abovedisplayskip{0.15cm}
\setlength\belowdisplayskip{0.15cm}
\boldsymbol{D}(n,i)=\frac{1}{N_R}\sum_{j=1}^{N_R}\left|\boldsymbol{M}(n,i,j)\right|,
\quad n\in\{1,\dots,N_F\},\ i\in\{1,\dots,N_D\}.
\end{equation}
The RTM captures coarse displacement dynamics by aggregating range responses over the Doppler axis:
\begin{equation}
\footnotesize
\setlength\abovedisplayskip{0.15cm}
\setlength\belowdisplayskip{0.15cm}
\boldsymbol{R}(n,j)=\frac{1}{N_D}\sum_{i=1}^{N_D}\left|\boldsymbol{M}(n,i,j)\right|,
\quad n\in\{1,\dots,N_F\},\ j\in\{1,\dots,N_R\}.
\end{equation}
The resulting $\boldsymbol{D}\in\mathbb{R}^{N_F\times N_D}$ and $\boldsymbol{R}\in\mathbb{R}^{N_F\times N_R}$ provide complementary velocity- and range-centric cues that serve as the basic radar representation for downstream tasks.

\begin{figure}[t]
    \centering
    \subfigure[Motion envelopes from radar and video.]{
        \includegraphics[width=0.58\linewidth]{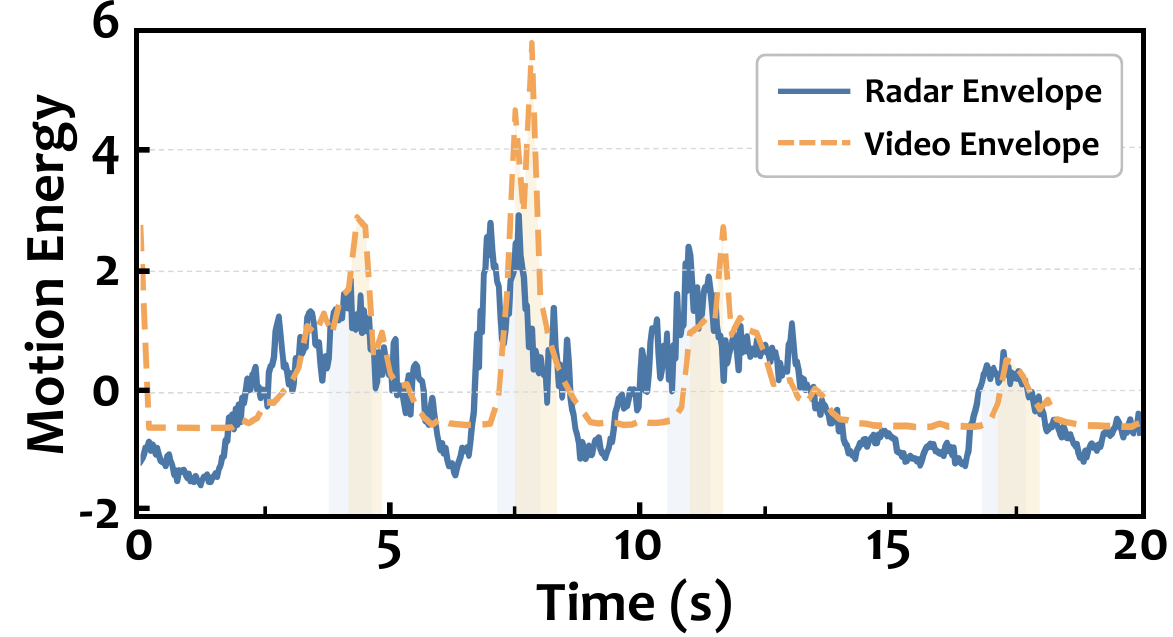}
        \label{fig:sync_raw}
    }
    \subfigure[Cross-correlation.]{
        \includegraphics[width=0.36\linewidth]{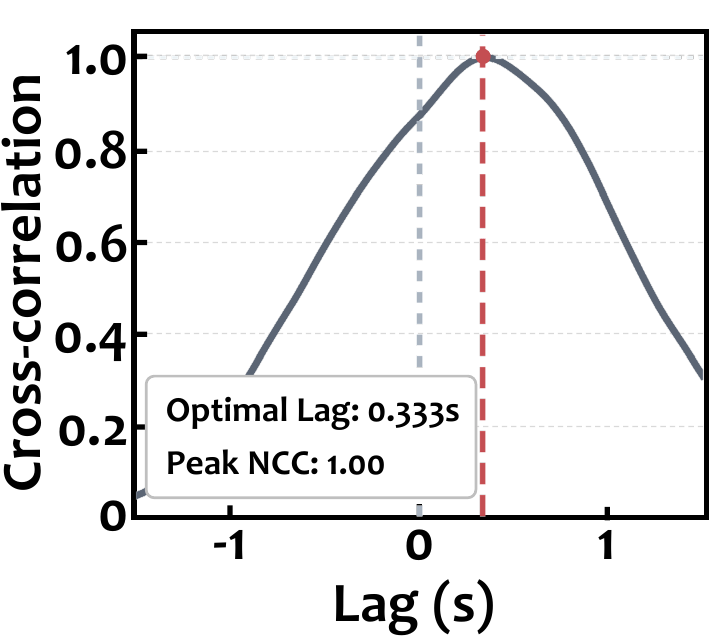}
        \label{fig:sync_ncc}
    }
    \vspace{-0.15in}
    \caption{\textbf{Radar-video temporal synchronization.} The optimal offset is estimated by maximizing the cross-correlation between the radar and video motion-energy envelopes.}
    \label{fig:sync}
    \vspace{-0.2in}
\end{figure}

\subsubsection{Temporal Synchronization and Segment Construction}
\label{subsubsec:sync}

To transfer visual semantics to paired radar observations, the synchronized RGB stream and mmWave signal must be temporally aligned before annotation. Because the radar and camera are triggered independently, we align them by maximizing the cross-correlation between their motion-energy envelopes.
We therefore adopt a motion-energy-based synchronization scheme that aligns the two modalities through their motion envelopes.

\begin{figure}[t!]
    \centering
    \includegraphics[width=0.98\linewidth]{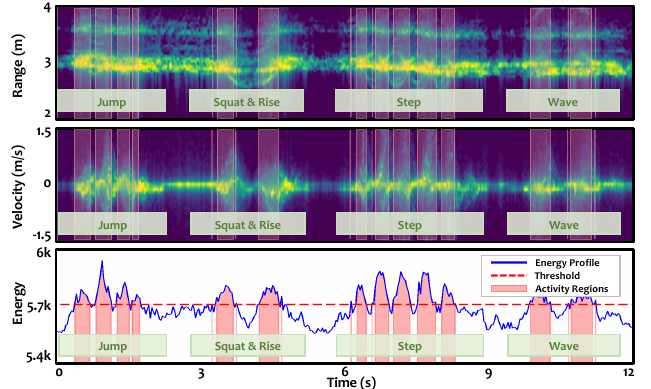}
    \vspace{-0.15in}
    \caption{\textbf{Activity segmentation from continuous mmWave streams.}}
    \label{fig:mmwave}
    \vspace{-0.2in}
\end{figure}

On the radar side, we derive a 1D motion-energy envelope from the DTM by frame-wise Doppler aggregation:
\begin{equation}
\footnotesize
\setlength\abovedisplayskip{0.15cm}
\setlength\belowdisplayskip{0.15cm}
E_r(t)=\sum_{i=1}^{N_D}\boldsymbol{D}(t,i), \quad t=1,\dots,T.
\end{equation}
On the video side, we estimate dense optical flow between consecutive RGB frames and summarize the per-frame flow magnitude into a visual motion-energy envelope:
\begin{equation}
\footnotesize
\setlength\abovedisplayskip{0.15cm}
\setlength\belowdisplayskip{0.15cm}
E_v(t)=\frac{1}{|\Omega|}\sum_{(x,y)\in\Omega}\left\|\mathbf{F}_t(x,y)\right\|_2,
\end{equation}
where $\mathbf{F}_t(x,y)$ denotes the optical-flow vector at pixel $(x,y)$ and $\Omega$ is the image domain. Because the two modalities have different sampling rates, we linearly resample the video envelope onto the radar timeline and normalize both signals to remove scale differences.

We then estimate the temporal offset by maximizing the cross-correlation between the radar and video envelopes:
\begin{equation}
\footnotesize
\setlength\abovedisplayskip{0.15cm}
\setlength\belowdisplayskip{0.15cm}
\begin{aligned}
C(\ell) &= \sum_t \hat{E}_v(t+\ell)\,\hat{E}_r(t),\\
\ell^\star &= \arg\max_{\ell} C(\ell),
\end{aligned}
\end{equation}
where $\hat{E}_r$ and $\hat{E}_v$ are the normalized radar and video motion-energy envelopes, and $\ell$ is the discrete lag. The optimal lag is converted into a time offset $\Delta t=\ell^\star/f_r$ using the radar frame rate $f_r$, and the video timestamps are shifted accordingly.

After synchronization, we partition the continuous radar stream into sample-level activity segments, so that each knowledge-base entry corresponds to one localized activity instance. Boundaries are detected from the radar motion-energy signal after suppressing quasi-static components near zero Doppler. An adaptive threshold generates a binary activity mask, which is further refined by smoothing, gap merging, boundary padding, and short-segment removal. The final boundaries are then applied consistently to both DTM and RTM for pseudo-label annotation, as illustrated in Fig.~\ref{fig:mmwave}.

\begin{figure}[t]
    \centering
    \includegraphics[width=0.99\linewidth]{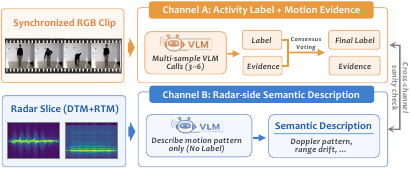}
    \vspace{-0.15in}
    \caption{\textbf{Dual-channel VLM annotation pipeline.}}
    \label{fig:vlmLabel}
    \vspace{-0.2in}
\end{figure}

\subsubsection{VLM-Supervised Semantic Annotations}
\label{subsubsec:vlm_semantic}

Given synchronized video--radar activity pairs, \SystemName assigns semantic metadata to each radar segment through a dual-channel cross-modal annotation pipeline, as shown in Fig.~\ref{fig:vlmLabel}. 
Each accepted knowledge-base entry finally stores four semantic fields: a pseudo-label, a structured motion evidence profile, a radar-side semantic description, and an acceptance status for reliability control.

\textbf{Primary Channel A: Video-side pseudo-labeling.}
The main annotation path operates on the synchronized RGB clip. Rather than accepting a single free-form response, we require the VLM to output exactly one label from the closed set together with a structured motion evidence profile covering displacement, cadence, arm action, torso action, and leg action. Each field is restricted to a predefined vocabulary so that obviously inconsistent outputs can be filtered deterministically. 
We query the VLM multiple times and count an output as a valid vote only when its predicted label is self-consistent with its structured evidence profile. The valid votes are then aggregated by adaptive early-stop voting to produce a candidate pseudo-label with consensus score $s_{\mathrm{ann}}$.

\begin{tcolorbox}[
  enhanced,
  colback=orange!2,
  colframe=orange!50!black,
  boxrule=0.6pt,
  arc=3pt,
  left=7pt,right=7pt,top=6pt,bottom=6pt,
  before skip=5pt, after skip=5pt,
  fontupper=\footnotesize,
  coltitle=black,
  title=\textbf{Channel A: Video-Side Pseudo-Labeling},
  colbacktitle=orange!12,
  attach boxed title to top left={xshift=2mm,yshift=-1.2mm},
  boxed title style={
    boxrule=0pt,
    arc=2pt,
    left=5pt,right=5pt,top=1.5pt,bottom=1.5pt
  }
]
\setlength{\parskip}{2pt}
\setlength{\parindent}{0pt}

\textbf{Input:} Synchronized RGB activity clip. \par
\textbf{Objective:} Produce exactly one candidate label from the closed set $\mathcal{C}$ together with a structured five-field motion evidence profile. \par
\textbf{Validation:} Retain only self-consistent outputs and aggregate valid votes through adaptive early-stop voting. \par
\textbf{Stored Output:} Candidate pseudo-label, consensus score $s_{\mathrm{ann}}$, representative evidence profile, and acceptance status.
\end{tcolorbox}

\textbf{Auxiliary Channel B: Radar-side semantic sanity check.}
We feed the paired DTM and RTM to a VLM under a blind prompting protocol in which the activity label is hidden. 
Instead of asking for classification, the model describes only observable motion characteristics, such as bursty versus periodic Doppler patterns, symmetric versus directional velocity structure, and stationary versus drifting range behavior. 
This channel does not produce the pseudo-label; it serves only as auxiliary semantic metadata and a coarse sanity check on the candidate label from the video side primary channel.

\begin{tcolorbox}[
  enhanced,
  colback=cyan!2,
  colframe=cyan!50!black,
  boxrule=0.6pt,
  arc=3pt,
  left=7pt,right=7pt,top=6pt,bottom=6pt,
  before skip=5pt, after skip=5pt,
  fontupper=\footnotesize,
  coltitle=black,
  title=\textbf{Channel B: Radar-Side Semantic Sanity Check},
  colbacktitle=cyan!12,
  attach boxed title to top left={xshift=2mm,yshift=-1.2mm},
  boxed title style={
    boxrule=0pt,
    arc=2pt,
    left=5pt,right=5pt,top=1.5pt,bottom=1.5pt
  }
]
\setlength{\parskip}{2pt}
\setlength{\parindent}{0pt}

\textbf{Input:} Paired DTM and RTM (activity label hidden). \par
\textbf{Objective:} Describe only observable motion characteristics, including temporal pattern, velocity structure, and range behavior, without predicting the activity label. \par
\textbf{Use:} Provide auxiliary semantic metadata and coarse motion cues for cross-channel compatibility checking. \par
\textbf{Stored Output:} One concise technical description together with coarse semantic cues for temporal pattern and range motion.
\end{tcolorbox}

The final pseudo-label is accepted only when the candidate label from Channel A satisfies two conditions: 
(1) it reaches sufficient consensus among valid video-side votes, and 
(2) it is not contradicted by the radar-side semantic description from Channel B. 
For the second condition, we extract coarse motion cues from the radar description and use simple compatibility checks to filter out pseudo-labels that contradict the observed radar pattern. 
As a result, unreliable cross-modal transfers are removed before entering the reusable knowledge base, improving the reliability of the stored semantic supervision for downstream retrieval and reasoning.
Importantly, Channel B is used only during offline knowledge-base curation and does not directly participate in deployment-time label resolution.

\subsection{Physics-Aware Numerical Retrieval}
\label{sec:retrieval}

To support deployment-time inference with physically comparable precedents, \SystemName retrieves historical radar samples in an explicit numerical feature space derived from DTM/RTM. 
Rather than relying on learned latent similarity, which can drift under changes in clutter, multipath, sensing geometry, or execution style, this design represents each segment by explicit motion descriptors that summarize its kinematic structure, such as displacement, cadence, spectral spread, and temporal dynamics. 
As a result, retrieval is grounded in motion-relevant physical attributes that are more stable and interpretable across domains, and returns a compact set of precedents for downstream reasoning.

\subsubsection{Physics-Driven Numerical Feature Space}
\label{subsec:physics_features}

For each radar segment, we extract an explicit physics-driven feature vector $\mathbf{x}\in\mathbb{R}^{D}$ from its DTM/RTM, so that retrieval similarity is defined by motion-relevant physical attributes rather than raw spectral appearance. The goal is not to encode every local detail of the spectrogram, but to summarize the segment in terms of kinematic properties that remain comparable across subjects and environments.

Specifically, for each radar segment, we construct a $D$-dimensional feature vector from its DTM/RTM. The vector aggregates three complementary physical perspectives. First, \textbf{kinematic and temporal features} characterize motion intensity and temporal dynamics, such as energy variation, duration, and periodicity, helping distinguish stationary, bursty, and sustained activities. Second, \textbf{micro-Doppler morphological features} describe the spectral structure of velocity signatures, including Doppler bandwidth, signed energy distribution, spectral spread, and symmetry, which capture differences in motion pattern and limb articulation. Third, \textbf{higher-order statistical features} summarize texture complexity and transient characteristics, such as spectral entropy, kurtosis, and SVD-based energy concentration, improving sensitivity to fine-grained differences while remaining robust to local noise. Through this design, $\mathbf{x}$ serves as a compact and interpretable physical representation for numerical retrieval.

Because different feature dimensions have heterogeneous units and value ranges, we apply per-dimension standardization using knowledge-base statistics so that the distance metric is not dominated by a small number of large-magnitude features. 
As a result, each radar segment corresponds to a standardized physical feature vector in a unified retrieval space, which provides the basis for subsequent subspace selection and nearest-neighbor retrieval.

\begin{table}[t]
\centering
\footnotesize
\renewcommand{\arraystretch}{1.0}
\setlength{\tabcolsep}{4pt}
\caption{Representative physics-driven features for retrieval.}
\vspace{-0.1in}
\label{tab:compact_features}
\resizebox{\columnwidth}{!}{
\begin{tabular}{l l p{0.72\linewidth}}
\toprule[1.5pt]
\textbf{Src.} & \textbf{Feature} & \textbf{Meaning} \\
\midrule
\multirow{5}{*}{\textbf{DTM}}
& Doppler BW (max/std) & Max and variation of Doppler bandwidth over time. \\
& Torso--Limb Ratio & Energy in torso-centered band vs.\ peripheral bins. \\
& Cadence (freq/strength) & Dominant motion cadence and its relative strength. \\
& Texture (contrast/energy) & GLCM texture of normalized DTM. \\
\midrule
\multirow{4}{*}{\textbf{RTM}}
& Total Displacement & Net range displacement from the tracked trajectory. \\
& Trajectory Linearity & $R^2$ of range-time linear fit (smooth drift vs.\ jitter). \\
& Mean Velocity & Mean absolute velocity from range track. \\
& Mean Acceleration & Mean absolute acceleration from range track. \\
\bottomrule[1.5pt]
\end{tabular}}
\vspace{-0.3in}
\end{table}

\subsubsection{ANOVA-Based Subspace Optimization and Nearest-Neighbor Retrieval}
\label{subsec:anova_retrieval}

Although the physics-driven feature space is explicit and interpretable, not all dimensions contribute equally to activity discrimination. Measuring distances in the full space may introduce redundant or weakly relevant features, which can dilute retrieval quality and reduce robustness under cross-environment variation. 
We therefore perform ANOVA~\cite{st1989analysis}-based feature ranking on the knowledge base and retain the most discriminative dimensions to form an optimized retrieval subspace.

\begin{figure}[t]
    \centering
    \subfigure[Top-ranked features.]{
        \includegraphics[width=0.47\linewidth]{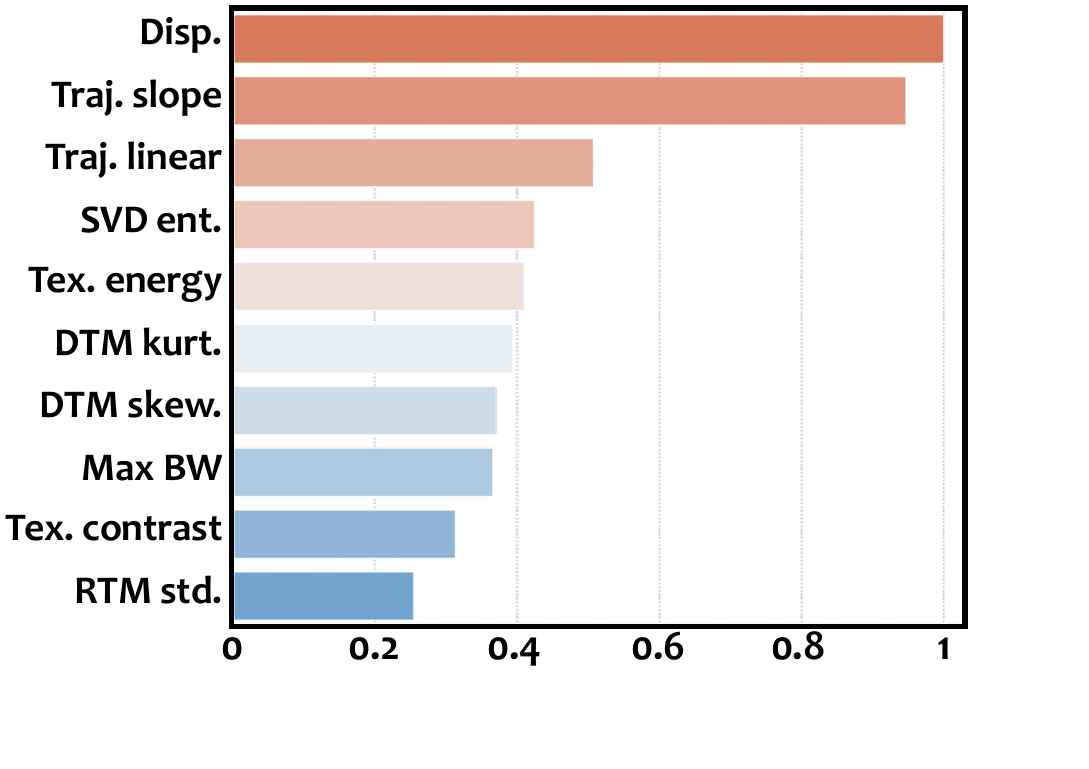}
        \label{fig:feature_importance_sub}
    }
    \subfigure[Class-feature response map.]{
        \includegraphics[width=0.47\linewidth]{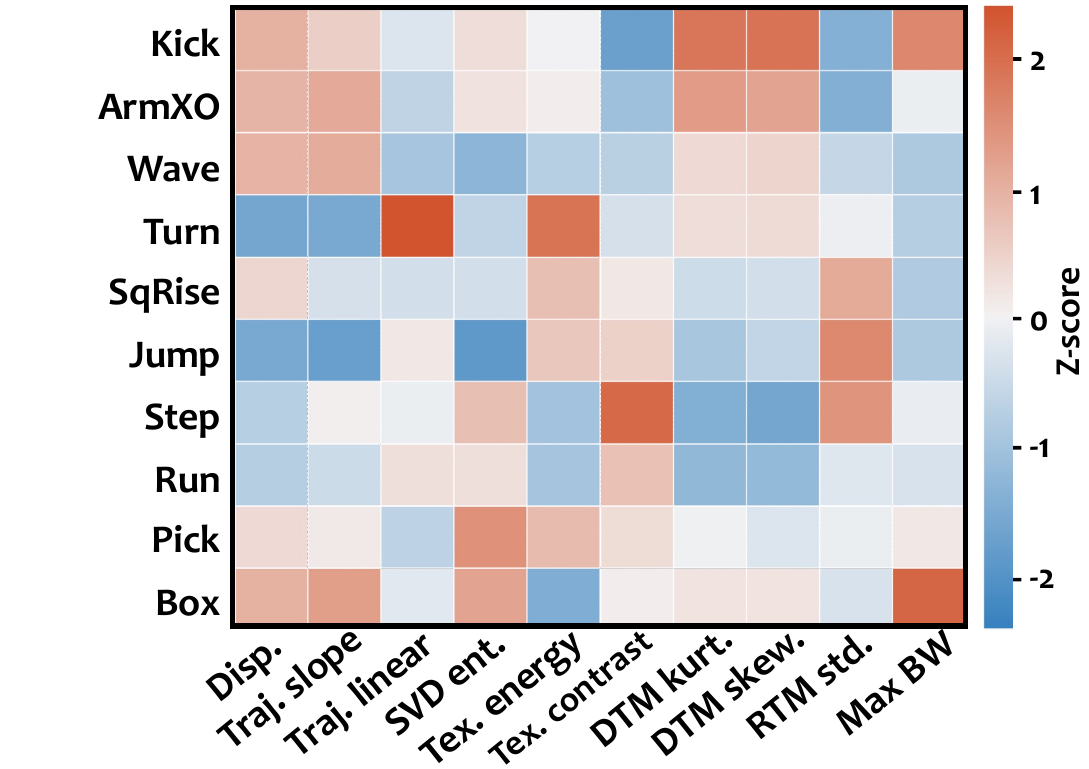}
        \label{fig:feature_heatmap_sub}
    }
    \vspace{-0.1in}
    \caption{\textbf{ANOVA-selected feature analysis.}
    (a) Normalized ANOVA $F$-scores of the top-ranked physics-driven features.
    (b) Class-wise mean z-score response patterns of representative activity classes.}
    \label{fig:feature_analysis}
\end{figure}

Specifically, for each feature dimension $x_j$, we compute the ANOVA $F$-statistic
\begin{equation}
\footnotesize
\label{eq:anova_f}
\setlength\abovedisplayskip{0.15cm}
\setlength\belowdisplayskip{0.15cm}
F_j=\frac{\sigma^2_{\mathrm{between}}(x_j)}{\sigma^2_{\mathrm{within}}(x_j)+\epsilon},
\end{equation}
where $\sigma^2_{\mathrm{between}}(x_j)$ and $\sigma^2_{\mathrm{within}}(x_j)$ denote the between-class and within-class variances, respectively, and $\epsilon$ is a small constant for numerical stability. We then rank all dimensions by $F_j$ and select the top-$K$ features to form the optimized subspace $\mathcal{S}_K$. Fig.~\ref{fig:feature_analysis} shows the resulting discriminative structure, where Doppler bandwidth, cadence, and range displacement consistently receive high scores.

Given a query sample $q$, we then retrieve a compact set of physically similar historical samples from the knowledge base $\mathcal{B}$ by top-$M$ nearest-neighbor search in $\mathcal{S}_K$:
\begin{equation}
\label{eq:knn_retrieval}
\footnotesize
\setlength\abovedisplayskip{0.15cm}
\setlength\belowdisplayskip{0.15cm}
\mathcal{N}_M(q)=\operatorname{TopM}_{d\in\mathcal{B}}
\left\|
\mathbf{x}^{(K)}_{q}-\mathbf{x}^{(K)}_{d}
\right\|_2 .
\end{equation}
The neighbor set $\mathcal{N}_M(q)$ serves as the retrieval evidence for downstream Council-of-Experts reasoning.

\subsection{Hybrid Expert Reasoning and Evolution}
\label{sec:reasoning}

\SystemName resolves each query through a structured \emph{Council of Experts} that consumes three complementary evidence sources: retrieval evidence from historical precedents, semantic evidence from the query spectrograms, and physical evidence from explicit kinematic descriptors. 
Rather than allowing a single free-form model call to interpret all inputs at once, \SystemName decomposes inference into role-specialized judgments and resolves the final label through physics-gated arbitration. 
This design provides a structured and auditable way to coordinate heterogeneous evidence sources, reducing unsupported semantic guesses and helping resolve semantic observations under retrieval support and physical feasibility.

\subsubsection{Council-of-Experts Reasoning}
\label{subsec:council_reasoning}

\begin{figure}[t]
    \centering
    \includegraphics[width=0.99\linewidth]{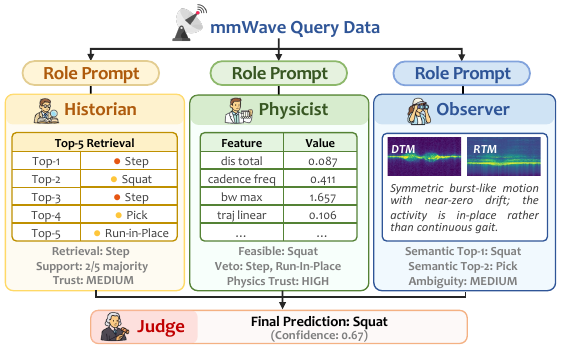}
    \caption{\textbf{Council-of-Experts reasoning.} Historian, Observer, and Physicist produce role-specific reports from retrieval, semantic, and physical evidence, and the Judge resolves the final activity label under physics-gated arbitration.}
    \label{fig:council_reasoning}
    \vspace{-0.15in}
\end{figure}

At inference time, the council operates on three complementary evidence sources for a query segment $q$: \textbf{retrieval evidence} from the physics-driven neighbor set, \textbf{physical evidence} from explicit kinematic descriptors, and \textbf{semantic evidence} from the query DTM/RTM. 
Rather than merging these cues through a single free-form LLM call, \SystemName resolves them through four role-specialized agents, as illustrated in Fig.~\ref{fig:council_reasoning}. 

The \textbf{Historian} summarizes retrieved precedents into a retrieval prior, the \textbf{Physicist} returns physically feasible and vetoed labels, the \textbf{Observer} produces a small set of semantic hypotheses with an ambiguity assessment, and the \textbf{Judge} resolves the final label through physics-first arbitration. 
In this design, semantic observations can refine the decision, but cannot override hard physical inconsistency.

Given a query segment $q$, the Historian converts the retrieved top-$M$ neighbors $\mathcal{N}_M(q)$ into a distance-weighted retrieval prior. For each neighbor $d\in\mathcal{N}_M(q)$ with label $y_d$, we assign a distance-based weight
\begin{equation}
\footnotesize
\setlength\abovedisplayskip{0.15cm}
\setlength\belowdisplayskip{0.15cm}
w_d=\frac{1}{\left\|\mathbf{x}^{(K)}_q-\mathbf{x}^{(K)}_d\right\|_2+\epsilon},
\end{equation}
where $\mathbf{x}^{(K)}$ denotes the physics-driven feature vector restricted to the ANOVA-selected Top-$K$ dimensions, and $\epsilon$ is a small constant for numerical stability. The label-wise retrieval support is then computed as
\begin{equation}
\footnotesize
\setlength\abovedisplayskip{0.15cm}
\setlength\belowdisplayskip{0.15cm}
\pi(c)=
\frac{\sum_{d\in\mathcal{N}_M(q)}\mathbb{I}[y_d=c]\cdot w_d}
{\sum_{d\in\mathcal{N}_M(q)} w_d},
\quad c\in\mathcal{C},
\label{eq:historian_support}
\end{equation}
where $\mathbb{I}[y_d=c]$ is the indicator function and $\pi(c)$ denotes the normalized retrieval support assigned to class $c$. The class with the largest $\pi(c)$ is treated as the leading hypothesis.

In parallel, the Physicist applies a small set of interpretable rule-based checks over the query's explicit physical descriptors and returns a feasible set and a vetoed set of labels. These checks are designed to rule out physically incompatible candidates rather than to classify the activity on their own. For example, locomotion classes require non-trivial directional displacement, near-stationary repetitive motions should exhibit limited range drift, and impulsive actions such as jumping or kicking are expected to produce short-duration burst-like Doppler responses with larger spectral spread.
Meanwhile, the Observer inspects only the query DTM/RTM and returns a small set of semantic hypotheses together with an ambiguity level.

The Judge performs final label resolution through a fixed arbitration order. It first removes vetoed labels and retains only physically feasible candidates. It then checks whether the Observer's leading semantic hypotheses agree with this feasible set. 
If multiple feasible candidates remain, the Judge selects the one with the largest retrieval support $\pi(c)$ from the Historian. 
If no stable semantic agreement exists, it backs off to the strongest retrieval-supported label within the feasible set. After prediction, we compute a post-hoc confidence score from retrieval strength, retrieval margin, and cross-role consistency. This score serves only as a reliability estimate and does not affect the Judge's final decision.

\begin{algorithm}[t]
\caption{\footnotesize{Offline zero-gradient protocol refinement}}
\label{alg:self_evolve}
\footnotesize
\KwIn{Development split $\mathcal{D}$, initial protocol $\mathcal{P}_0$, iterations $T$}
\KwOut{Best protocol $\mathcal{P}_{\mathrm{best}}$}

$\mathcal{P}_{\mathrm{best}} \leftarrow \mathcal{P}_0$, \quad
$s_{\mathrm{best}} \leftarrow -\infty$\;

\For{$t=1$ \KwTo $T$}{
  Evaluate $\mathcal{P}_{t-1}$ on $\mathcal{D}$ to obtain score $s_t$ and traces $\mathcal{T}_t$ 
  
  \If{$s_t > s_{\mathrm{best}}$}{
    $\mathcal{P}_{\mathrm{best}} \leftarrow \mathcal{P}_{t-1}$\;
    $s_{\mathrm{best}} \leftarrow s_t$ \tcp*{update best protocol}
  }

  Split $\mathcal{T}_t$ into success traces $\mathcal{T}_t^{+}$ and failure traces $\mathcal{T}_t^{-}$\;
  
$\tilde{\mathcal{P}} \leftarrow \mathrm{ReviseProtocol}(\mathcal{P}_{t-1}, \mathcal{T}_t^{-}, \mathcal{T}_t^{+})$\;
$\mathcal{P}_t \leftarrow \tilde{\mathcal{P}}$\;
\If{$\mathrm{Score}(\tilde{\mathcal{P}}, \mathcal{D}) < s_{\mathrm{best}} - \delta$}{
  $\mathcal{P}_t \leftarrow \mathcal{P}_{\mathrm{best}}$ \tcp*{rollback for stability}
}
}
\Return{$\mathcal{P}_{\mathrm{best}}$}\;
\end{algorithm}

\subsubsection{Zero-Gradient Self-Evolution}
\label{subsubsec:self_evolution}

To improve the stability of Council-of-Experts reasoning, \SystemName further refines its reasoning protocol through an offline zero-gradient self-evolution procedure. This step does not modify the underlying LLM, perform backpropagation, or introduce target-domain adaptation; it only updates the protocol text that governs how retrieval, semantic, and physical evidence are coordinated during reasoning.

Specifically, we maintain an editable protocol template $\mathcal{P}$ and evaluate it on a held-out development split $\mathcal{D}$. At each iteration, the current protocol is executed together with the retrieved evidence, semantic observations, and physical constraints to obtain predictions and role-specific diagnostic traces. These traces are then divided automatically into success and failure cases according to whether the predicted label matches the reference label on $\mathcal{D}$.
A second LLM is then used to revise only the protocol section based on sampled success and failure traces. The revision is instructed to improve failure cases while preserving effective reasoning patterns from successful ones. 
To avoid unstable drift, we keep track of the best-performing protocol across iterations and roll back whenever a revised protocol causes a significant performance degradation, as summarized in Alg.~\ref{alg:self_evolve}.

This procedure is performed only once offline and refines how the council arbitrates among heterogeneous evidence sources, rather than changing the model itself. As a result, \SystemName improves reasoning consistency while preserving the deployment-time training-free nature of inference.

\section{Evaluation}
\label{sec:evaluation}

\subsection{Experimental Methodology}

\subsubsection{Implementation Details}
\begin{figure}[t]
\centering
\includegraphics[width=0.99\linewidth]{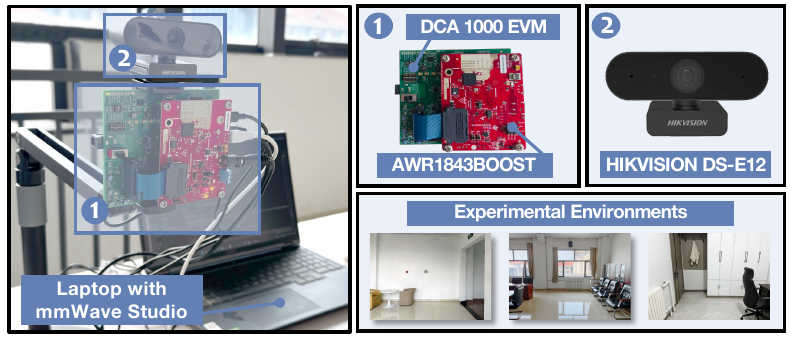}
\caption{\textbf{Experimental setup.} A laptop-controlled TI AWR1843BOOST with DCA1000EVM is used for radar acquisition, while the RGB camera is used only for offline knowledge-base construction in three indoor environments.}
\label{fig:exp_setup}
\vspace{-0.15in}
\end{figure}

We implement \SystemName on a commodity TI AWR1843BOOST mmWave radar~\cite{TI_AWR1843} operating at 77\,GHz, with a DCA1000EVM~\cite{TI_DCA1000} for raw radar data capture, and a synchronized HIKVISION DS-E12 RGB camera~\cite{Hikv} used only during offline cross-modal knowledge-base construction. 
As shown in Fig.~\ref{fig:exp_setup}, a laptop running mmWave Studio is used to control data acquisition.

\subsubsection{Data Collection}
We evaluate \SystemName on a self-collected mmWave HAR dataset\footnote{Ethical approval has been obtained from the corresponding organization.} acquired from 8 participants (5 male and 3 female) in 3 indoor environments: a corridor, a meeting room, and a bedroom (Fig.~\ref{fig:exp_setup}). Each participant performs 12 predefined activities in each environment, with 5--10 repetitions per activity. After preprocessing and motion-energy-based segmentation (Sec.~\ref{subsubsec:sync}), the dataset contains 2,568 radar segments. Each segment is represented by its DTM/RTM, a 25-dimensional physics-driven feature vector, and the associated semantic metadata produced during offline knowledge-base construction. Reference labels are determined by the predefined activity script and manually verified when necessary. 
The dataset is partitioned into three disjoint subsets for knowledge-base construction, offline protocol refinement, and final testing, with no overlap. No sample is shared across subsets, and the final test set is kept isolated from offline protocol refinement.

\begin{figure}[t]
    \centering
    \includegraphics[width=0.92\linewidth]{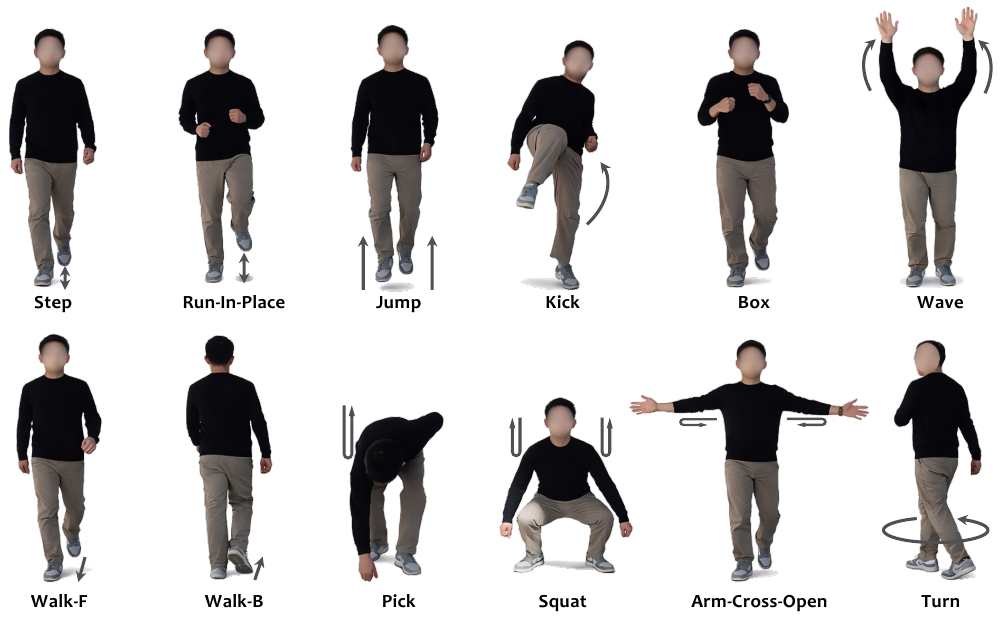}
    \caption{\textbf{Activities in the dataset.} 2,568 radar segments are collected from 8 participants performing 12 activities across 3 environments.}
    \label{fig:activity_class}
    \vspace{-0.25in}
\end{figure}

\subsubsection{Metrics}
We evaluate \SystemName from three main perspectives. 
\textbf{(1) Overall Accuracy} measures the fraction of correctly classified test samples and is used as the primary indicator of end-to-end recognition performance. 
\textbf{(2) Macro-F1} computes the unweighted mean of per-class F1 scores and reflects balanced performance across activity categories, especially under class-specific confusion. 
\textbf{(3) VLM Annotation Accuracy} evaluates the quality of offline pseudo-label generation by comparing accepted annotations against manually verified reference labels. 
In addition, we further analyze confusion patterns for fine-grained class-level behavior and examine the relationship between post-hoc confidence and prediction correctness to assess reliability.

\begin{figure*}[t]
\centering
\subfigure[Overall accuracy.]{
\begin{minipage}[t]{0.24\linewidth}
\centering
\includegraphics[width=0.95\linewidth]{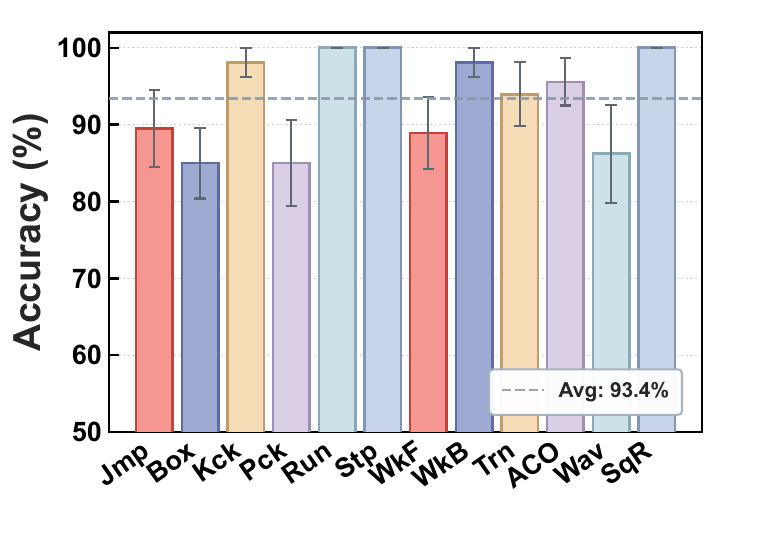}
\label{fig:eval_overall_acc}
\vspace{-5mm}
\end{minipage}%
}%
\subfigure[Confusion matrix.]{
\begin{minipage}[t]{0.24\linewidth}
\centering
\includegraphics[width=0.95\linewidth]{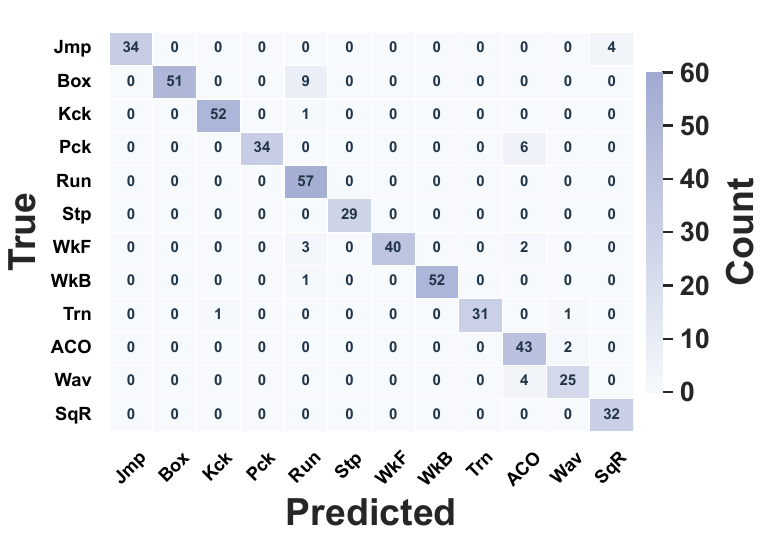}
\label{fig:eval_confusion}
\vspace{-5mm}
\end{minipage}%
}%
\subfigure[Confidence vs.\ prediction accuracy.]{
\begin{minipage}[t]{0.24\linewidth}
\centering
\includegraphics[width=0.95\linewidth]{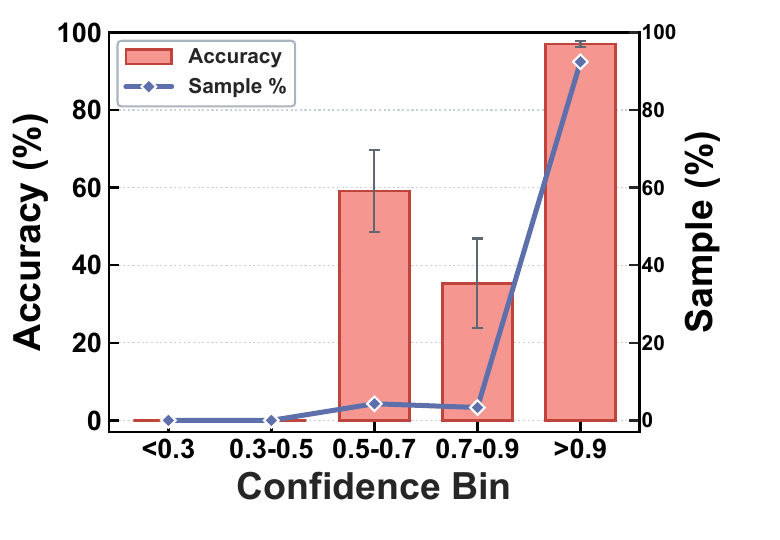}
\label{fig:eval_confidence}
\vspace{-5mm}
\end{minipage}%
}%
\subfigure[VLM annotation accuracy.]{
\begin{minipage}[t]{0.24\linewidth}
\centering
\includegraphics[width=0.95\linewidth]{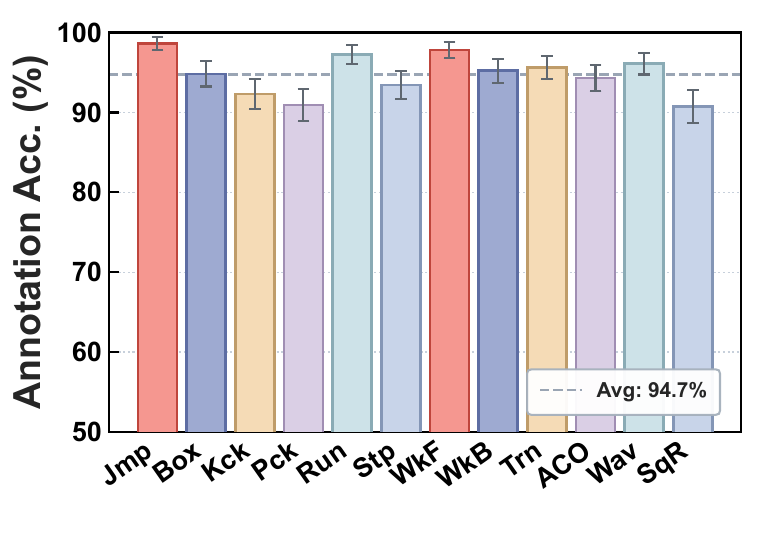}
\label{fig:eval_vlm_anno}
\vspace{-5mm}
\end{minipage}%
}%
\vspace{-5mm}
\centering
\caption{\textbf{Overall performance. \SystemName achieves 93.39\% accuracy without per-domain retraining or target-domain adaptation.}}
\label{fig:eval_overall}
\vspace{-0.1in}
\end{figure*}

\subsection{Overall Performance}

\subsubsection{Recognition Performance}
\label{subsec:har_performance}

We first evaluate the end-to-end recognition performance of \SystemName. 
As shown in Fig.~\ref{fig:eval_overall_acc}, \SystemName achieves 93.39\% overall accuracy and 93.61\% macro-F1 without per-domain retraining or target-domain adaptation. These results indicate that evidence-grounded inference over reusable radar knowledge can achieve strong recognition performance even without fitting a task-specific classifier.
The confusion matrix in Fig.~\ref{fig:eval_confusion} shows that the remaining errors are concentrated mainly among a few semantically related and physically similar activities, rather than being uniformly distributed across the label space. 
This pattern is consistent with the \SystemName design: retrieval provides a precedent-grounded prior, physical constraints rule out incompatible candidates, and semantic observations help resolve residual ambiguity when physical similarity alone is insufficient.

We further examine whether the post-hoc confidence score $s$ reflects prediction reliability. 
As shown in Fig.~\ref{fig:eval_confidence}, most samples fall into the high-confidence region ($s>0.9$), where prediction accuracy approaches 98\%, whereas medium- and low-confidence cases are noticeably less reliable. This result suggests that the confidence score provides a useful reliability signal and may support selective prediction in practical deployment.

\subsubsection{VLM Annotation Quality}
\label{subsec:vlm_annotation}

We then evaluate the quality of the offline VLM annotation pipeline used for knowledge-base construction. 
The structured evidence-gated consensus achieves an overall annotation accuracy of 94.7\%.  
Among all clips, 85.3\% are marked as strong accept ($s_{\mathrm{ann}} \geq 0.8$), while 4.7\% are rejected due to low consensus.
These results indicate that the primary video-side annotation channel already provides reliable pseudo-labels for most activity segments.

As shown in Fig.~\ref{fig:eval_vlm_anno}, the remaining errors follow an intuitive pattern. Activities with distinctive visual signatures achieve near-ceiling annotation accuracy, whereas visually similar classes, such as \emph{Picking} and \emph{Squatting and Rising}, are more challenging. This confirms that the main source of annotation error lies in semantic similarity between activities rather than random label noise.
The radar-side sanity check further filters out 3.1\% of physically incompatible pseudo-labels before they are written into the knowledge base. Overall, these results suggest that the accepted knowledge-base entries provide sufficiently reliable semantic supervision for downstream retrieval and council reasoning.

\begin{table}[t]
\centering
\footnotesize
\caption{Performance comparison with baselines.}
\vspace{-0.1in}
\label{tab:sota_performance}
\resizebox{\columnwidth}{!}{
\begin{tabular}{l l|cc|c}
\toprule[1.5pt]
\textbf{Category} & \textbf{Method} & \textbf{Acc. (\%)} & \textbf{Macro-F1 (\%)} & \textbf{Train-free} \\
\midrule
\multirow{2}{*}{\textbf{Inference-only}}
& Retrieval-Only        & \tablebar{100}{65.00} & \tablebar{100}{61.64} & \CIRCLE \\
& LLM-Direct      & \tablebar{100}{13.10} & \tablebar{100}{13.85} & \CIRCLE \\
\midrule
\multirow{3}{*}{\textbf{Supervised}}
& CNN                   & \tablebar{100}{91.76} & \tablebar{100}{90.03} & \Circle \\
& CNN+BiLSTM            & \tablebar{100}{88.82} & \tablebar{100}{85.50} & \Circle \\
& ViT                   & \tablebar{100}{94.12} & \tablebar{100}{92.47} & \Circle \\
\midrule
\multirow{3}{*}{\textbf{Adapted baselines}}
& mmCLIP-light          & \tablebar{100}{87.06} & \tablebar{100}{81.70} & \Circle \\
& SynMotion-light & \tablebar{100}{72.35} & \tablebar{100}{70.68} & \Circle \\
& EI-adapt      & \tablebar{100}{68.00} & \tablebar{100}{63.99} & \Circle \\
\midrule\midrule
\textbf{Ours}
& \textbf{\SystemName}  & \textbf{\tablebar[blue!25]{100}{93.39}} & \textbf{\tablebar[blue!25]{100}{93.61}} & \CIRCLE \\
\bottomrule[1.5pt]
\end{tabular}
}
\raggedright
{\scriptsize \textit{Note:} Adapted baselines follow a unified protocol and are not official reproductions.}
\vspace{-6mm}
\end{table}

\subsection{Comparison with Baselines}

Tab.~\ref{tab:sota_performance} reports a contextual comparison between \SystemName and three groups of methods: inference-only methods, supervised models, and paper-inspired reference implementations.

Among the inference-only baselines, Retrieval-Only achieves 65.00\% accuracy and 61.64\% macro-F1, indicating that physics-driven nearest-neighbor voting already provides a meaningful precedent prior, but is insufficient to resolve fine-grained ambiguity on its own. LLM-Direct performs substantially worse, achieving 13.10\% accuracy and 13.85\% macro-F1, which suggests that direct closed-set prediction from radar observations remains unreliable without retrieval grounding or explicit physical constraints.

Supervised models achieve strong performance after task-specific training. Among them, ViT performs best with 94.12\% accuracy and 92.47\% macro-F1, while CNN and CNN+BiLSTM achieve 91.76\% and 88.82\% accuracy, respectively. This result provides a useful upper-bound reference for a deployment-specific training regime.

We further include several paper-inspired reference implementations based on representative prior methods, including mmCLIP-light~\cite{cao2024mmclip}, SynMotion-light~\cite{zhang2022synthesized}, and EI-adapt~\cite{jiang2018towards}. Since official implementations are unavailable and the experimental settings differ from ours, these methods are instantiated under a unified DTM/RTM-based protocol following the methodological descriptions in the original papers. These results are intended as contextual reference rather than reproduction-level benchmarking.
Under this protocol, mmCLIP-light achieves 87.06\% accuracy, while SynMotion-light and EI-adapt perform lower.

Within this comparison framework, \SystemName achieves 93.39\% accuracy and 93.61\% macro-F1 without per-domain retraining or target-domain adaptation. 
Its performance is close to the strongest supervised baseline (ViT), while also outperforming the inference-only and adapted baselines under the same unified evaluation protocol.
Overall, these results show that combining physics-aware retrieval with multi-role reasoning can approach supervised performance while preserving deployment-time training-free inference.

\begin{figure*}[t]
\centering
\subfigure[Impact of knowledge base size.]{
\begin{minipage}[t]{0.24\linewidth}
\centering
\includegraphics[width=0.95\linewidth]{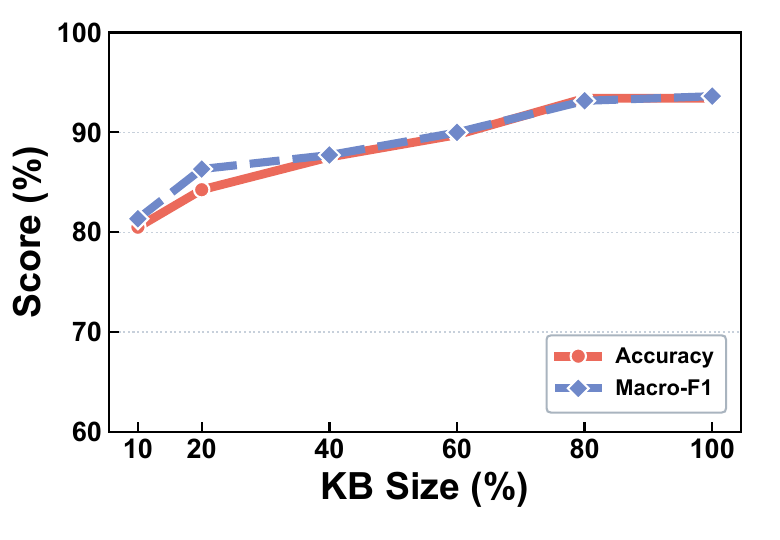}
\label{fig:eval_kb_size}
\vspace{-5mm}
\end{minipage}%
}%
\subfigure[Impact of number of categories.]{
\begin{minipage}[t]{0.24\linewidth}
\centering
\includegraphics[width=0.95\linewidth]{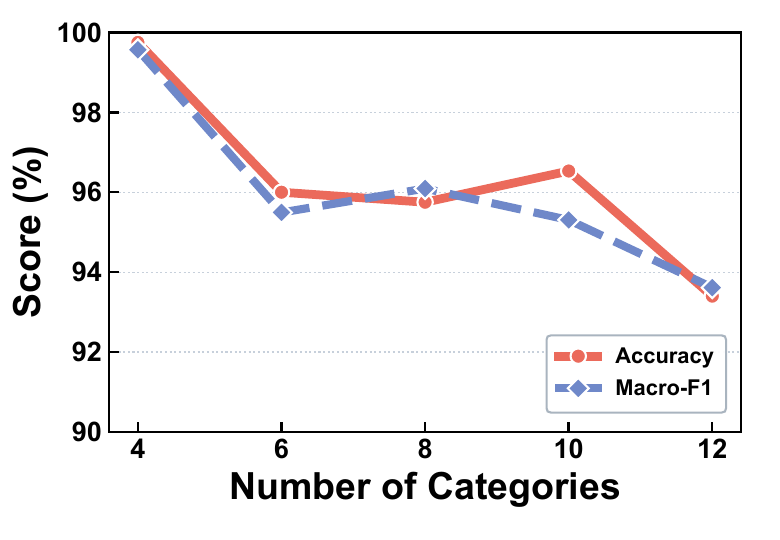}
\label{fig:eval_category}
\vspace{-5mm}
\end{minipage}%
}%
\subfigure[Impact of retrieval Top-$M$.]{
\begin{minipage}[t]{0.24\linewidth}
\centering
\includegraphics[width=0.95\linewidth]{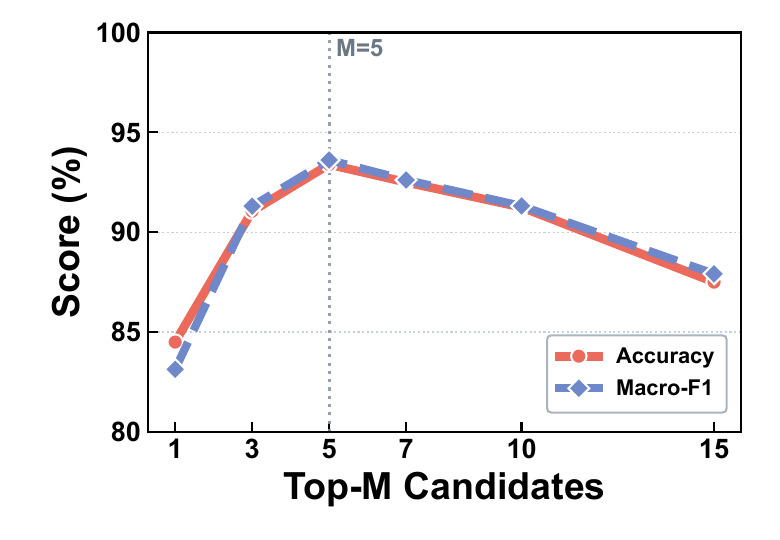}
\label{fig:eval_topM}
\vspace{-5mm}
\end{minipage}%
}%
\subfigure[ANOVA Top-$K$ sensitivity.]{
\begin{minipage}[t]{0.24\linewidth}
\centering
\includegraphics[width=0.95\linewidth]{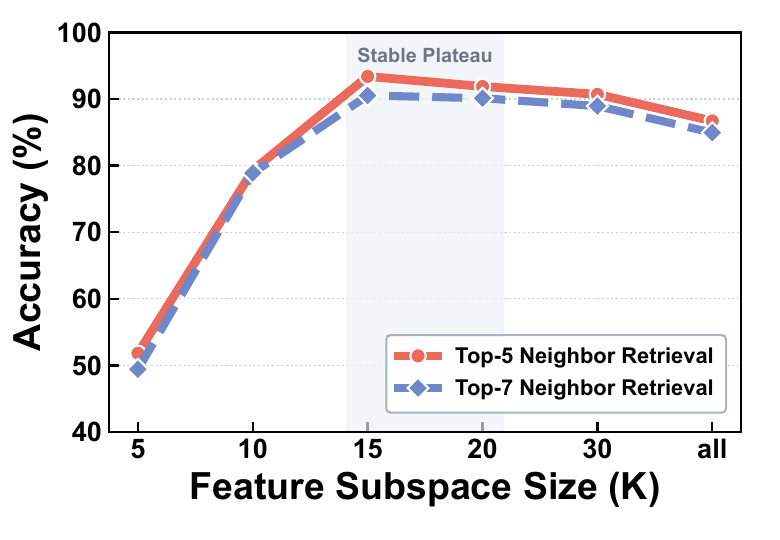}
\label{fig:eval_anova_sensitivity}
\vspace{-5mm}
\end{minipage}%
}%
\vspace{-5mm}
\centering
\caption{\textbf{Micro-benchmark results with respect to knowledge base size, category scale, retrieval depth, and feature subspace size.} }
\label{fig:eval_ablation_micro}
\vspace{-0.15in}
\end{figure*}

\subsection{Ablation Studies}
\label{subsec:ablation}

Tab.~\ref{tab:ablation} reports the ablation results of the major components in \SystemName. The full system achieves 93.39\% accuracy and 93.61\% macro-F1. Each ablated variant removes one component while keeping the others unchanged. Overall, the results show that retrieval grounding is the most influential component, while physical validation, semantic observation, and offline protocol refinement all make consistent contributions to the final performance.

\vspace{-0.05in}

\subsubsection{Impact of retrieval grounding (w/o Historian).}
Removing the Historian causes the largest degradation, reducing accuracy from 93.39\% to 73.60\% and macro-F1 from 93.61\% to 73.21\%. 
This result indicates that retrieval-based historical grounding provides the primary decision prior of the system. Without precedent support from the knowledge base, the council loses its strongest source of evidence and becomes substantially less reliable.

\subsubsection{Impact of physical validation (w/o Physicist).}
Removing the Physicist reduces performance to 83.33\% accuracy and 85.31\% macro-F1, a 10.06-point drop in accuracy. This indicates that physical validation is important for ruling out semantically plausible but quantitatively inconsistent candidates, and that retrieval support alone may not ensure physically valid decisions.

\subsubsection{Impact of semantic observation (w/o Observer).}
Without the Observer, performance drops to 86.90\% accuracy and 86.48\% macro-F1. Although the degradation is smaller than that of removing the Historian or Physicist, it shows that semantic observation provides useful complementary cues when physically similar precedents remain ambiguous.

\subsubsection{Impact of self-evolution (w/o Self-Evo).}
Disabling self-evolution reduces performance to 88.24\% accuracy and 87.81\% macro-F1. This indicates that offline protocol refinement improves how retrieval, semantic, and physical evidence are coordinated during council reasoning, even though it is not itself the primary source of recognition capability.

\begin{table}[t]
\centering
\footnotesize
\caption{Ablation results of major components in \SystemName.}
\vspace{-0.1in}
\label{tab:ablation}
\resizebox{\columnwidth}{!}{
\begin{tabular}{l|cc|cc}
\toprule[1.5pt]
\textbf{Variant} & \textbf{Acc. (\%)} & \textbf{Macro-F1 (\%)} & \textbf{$\Delta$Acc (\%)} & \textbf{$\Delta$Macro-F1 (\%)}\\
\midrule
\textbf{Full} & \tablebar[blue!25]{100}{93.39} & \tablebar[blue!25]{100}{93.61} & -- & -- \\
\textbf{w/o Historian} & \tablebar{100}{73.60} & \tablebar{100}{73.21} & \deltabar{25}{19.79} & \deltabar{25}{20.40} \\
\textbf{w/o Physicist} & \tablebar{100}{83.33} & \tablebar{100}{85.31} & \deltabar{25}{10.06} & \deltabar{25}{8.30} \\
\textbf{w/o Observer} & \tablebar{100}{86.90} & \tablebar{100}{86.48} & \deltabar{25}{6.49} & \deltabar{25}{7.13} \\
\textbf{w/o Self-Evo} & \tablebar{100}{88.24} & \tablebar{100}{87.81} & \deltabar{25}{5.15} & \deltabar{25}{5.80} \\
\bottomrule[1.5pt]
\end{tabular}}
\vspace{-0.2in}
\end{table}

\begin{figure*}[t]
\centering
\subfigure[Cross-LLM performance.]{
\begin{minipage}[t]{0.24\linewidth}
\centering
\includegraphics[width=0.95\linewidth]{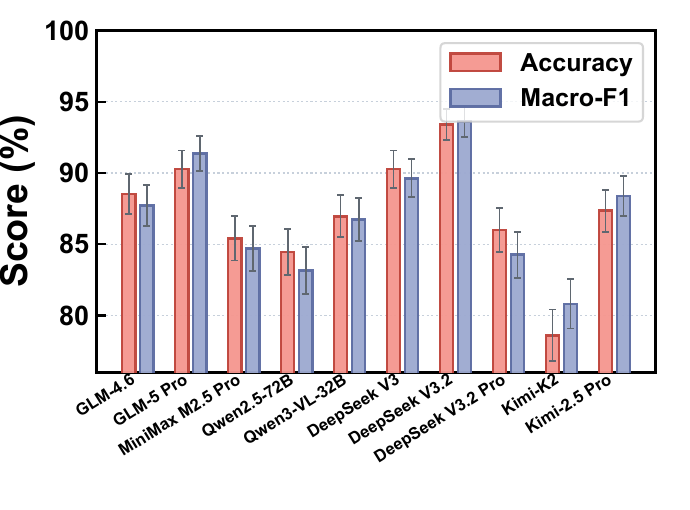}
\label{fig:eval_cross_llm}
\vspace{-5mm}
\end{minipage}%
}%
\subfigure[Cross-environment generalization.]{
\begin{minipage}[t]{0.24\linewidth}
\centering
\includegraphics[width=0.95\linewidth]{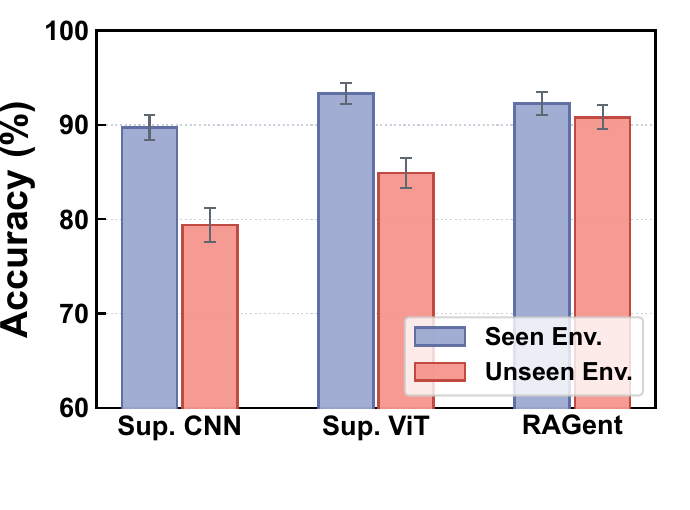}
\label{fig:eval_cross_env}
\vspace{-5mm}
\end{minipage}%
}%
\subfigure[Cross-subject generalization.]{
\begin{minipage}[t]{0.24\linewidth}
\centering
\includegraphics[width=0.95\linewidth]{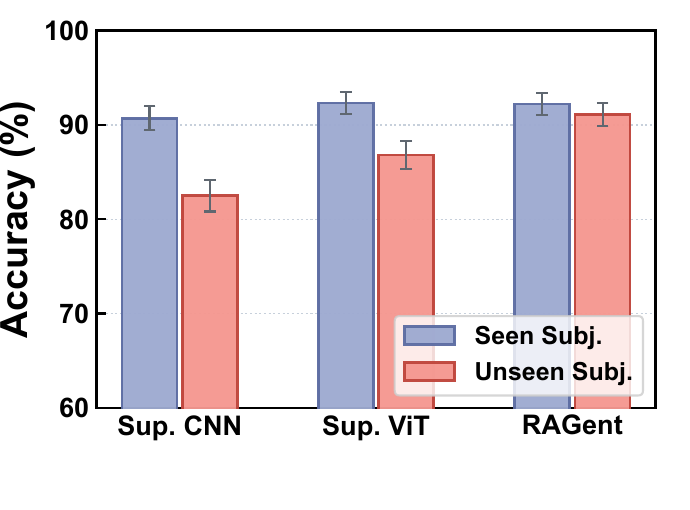}
\label{fig:eval_cross_subject}
\vspace{-5mm}
\end{minipage}%
}%
\subfigure[Cross-date generalization.]{
\begin{minipage}[t]{0.24\linewidth}
\centering
\includegraphics[width=0.95\linewidth]{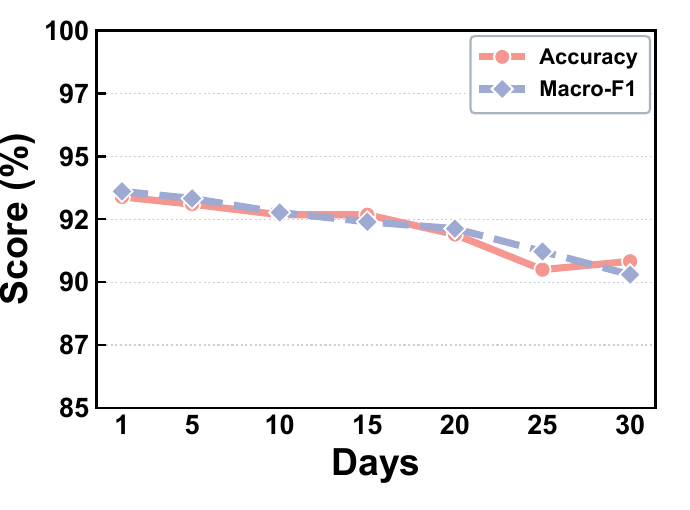}
\label{fig:eval_cross_date}
\vspace{-5mm}
\end{minipage}%
}%
\vspace{-5mm}
\centering
\caption{\textbf{Robustness and generalization of \SystemName} across different LLM backends and distribution-shift settings.}
\label{fig:eval_robustness}
\vspace{-0.15in}
\end{figure*}

\subsection{Micro-Benchmark Evaluation}
\label{subsec:micro_benchmark}

We then examine the sensitivity of \SystemName to several key factors, including knowledge base size, category scale, retrieval depth, and ANOVA-selected feature subspace size.

\subsubsection{Impact of Knowledge Base Size}
\label{subsec:kb_size}

Fig.~\ref{fig:eval_kb_size} shows the effect of varying the proportion of data used for knowledge-base construction.
As the size grows, performance improves steadily: accuracy increases from 80.53\% at 10\% KB to 93.39\% at full scale, while macro-F1 rises from 81.39\% to 93.61\%.
The gain becomes marginal after roughly 80\% coverage, suggesting that once sufficient historical diversity is accumulated, additional memory mainly provides diminishing returns.
Notably, even with only 10\% of the knowledge base, \SystemName still maintains over 80\% accuracy, indicating reasonable robustness under limited-memory conditions.

\subsubsection{Impact of Number of Categories}
\label{subsec:category_scale}

Fig.~\ref{fig:eval_category} evaluates scalability as the number of activity categories increases from 4 to 12.
\SystemName maintains strong performance throughout, with accuracies of 99.75\%, 96.00\%, 95.75\%, 96.53\%, and 93.39\% for 4, 6, 8, 10, and 12 classes, respectively.
Macro-F1 closely follows the same trend, indicating stable class-level behavior as the label space expands.
These results suggest that the framework scales gracefully, with performance degradation remaining limited even in the full 12-class setting.

\subsubsection{Impact of Retrieval Depth}
\label{subsec:topM}

Fig.~\ref{fig:eval_topM} shows the sensitivity to retrieval depth $M$.
Performance improves from 84.50\% at $M=1$ to a peak of 93.39\% at $M=5$, and then gradually declines to 92.50\% at $M=7$, 91.25\% at $M=10$, and 87.50\% at $M=15$.
This trend reflects a natural trade-off: overly small $M$ provides insufficient contextual evidence, whereas overly large $M$ introduces noisy or less relevant neighbors.
A moderate retrieval depth, especially around $M=5$--7, therefore provides the most stable operating point.

\subsubsection{Impact of Feature Subspace Size}
\label{subsec:topK}

Fig.~\ref{fig:eval_anova_sensitivity} evaluates the sensitivity to the ANOVA-selected feature subspace size $K$.
When $K$ is too small, the retrieval space is under-expressive: accuracy is only 51.82\% at $K=5$ and 79.29\% at $K=10$.
Performance peaks at $K=15$, reaching 93.39\% accuracy and 93.61\% macro-F1, remains stable over the range $K=15-20$, and then drops slightly to 86.71\% when all features are used.
This result indicates that a moderate ANOVA-optimized subspace achieves the best balance between discriminative power and robustness, while including too many weakly relevant features can dilute retrieval quality.

\subsection{Robustness and Generalization}
\label{subsec:robustness}

We evaluate the robustness of \SystemName across different LLM backends and under cross-environment, cross-subject, and cross-date settings. 
Overall, results show that \SystemName remains stable under multiple sources of variation, which is consistent with its design philosophy rather than deployment-specific model optimization.

\subsubsection{Cross-LLM Performance}
\label{subsec:cross_llm}

We first examine the sensitivity of \SystemName to the choice of LLM backend while keeping all other components unchanged. As shown in Fig.~\ref{fig:eval_cross_llm}, performance varies across models, but the variation remains moderate overall. The strongest backend (DeepSeek V3.2) reaches about 93\% in both accuracy and macro-F1, while mid-sized models remain in the 85\%--90\% range, and even smaller models stay close to 80\%. These results suggest that backend capability affects the performance ceiling, but the overall framework remains effective across different LLMs.

\subsubsection{Cross-Environment Generalization}
\label{subsec:cross_env}

We next evaluate environmental robustness by constructing the knowledge base from two environments and testing on the remaining one. 
As shown in Fig.~\ref{fig:eval_cross_env}, supervised baselines degrade noticeably in unseen environments, whereas \SystemName maintains about 91\% accuracy with only a modest drop. This result suggests that the explicit physics-driven retrieval space is less sensitive to environment-specific clutter than deployment-specific learned classifiers, and that the council can still resolve activity labels reliably under environmental shift.

\subsubsection{Cross-Subject Generalization}
\label{subsec:cross_subject}

We further evaluate robustness to subject variation using a leave-two-subjects-out protocol. As shown in Fig.~\ref{fig:eval_cross_subject}, \SystemName maintains accuracy above 91\% with relatively small fluctuations, while supervised baselines show more noticeable degradation on unseen subjects. Compared with the cross-environment setting, the gap here is smaller, suggesting that the kinematic structure of an activity remains more stable across subjects than across sensing environments. This result is consistent with the use of explicit motion descriptors and structured physics-aware reasoning instead of deployment-specific classifier fitting.

\subsubsection{Cross-Date Generalization}
\label{subsec:cross_date}

Finally, we evaluate robustness under temporal drift across different dates. As shown in Fig.~\ref{fig:eval_cross_date}, when the time gap increases from 1 day to 30 days, both accuracy and macro-F1 decline only slightly, within about 2--3 percentage points. Even with the longest gap, performance remains above 90\%. This suggests that the physics-driven retrieval space remains relatively stable over time, while structured council reasoning helps absorb the residual uncertainty introduced by temporal variation.

\begin{table}[t]
\centering
\footnotesize
\caption{Inference overhead of \SystemName.}
\vspace{-0.1in}
\label{tab:overhead}
\resizebox{\columnwidth}{!}{
\begin{tabular}{l|rrr}
\toprule[1.5pt]
\textbf{Stage} & \textbf{Latency} & \textbf{Tokens (K)} & \textbf{Cost (\$/1K samples)} \\
\midrule
\textbf{VLM-based semantic observation} & \tablebar[brown!12]{8}{1.800}\,s & \tablebar[gray!18]{3}{1.240} & \tablebar[blue!10]{1}{0.353} \\
\textbf{Physics-aware retrieval} & 0.735\,ms & -- & -- \\
\textbf{Agentic council reasoning} & \tablebar[brown!12]{8}{5.442}\,s & \tablebar[gray!18]{3}{1.743} & \tablebar[blue!10]{1}{0.485} \\
\midrule\midrule
\textbf{Total} & \textbf{\tablebar[brown!18]{8}{7.243}}\,\textbf{s} & \textbf{\tablebar[gray!25]{3}{2.983}} & \textbf{\tablebar[blue!16]{1}{0.838}} \\
\bottomrule[1.5pt]
\end{tabular}}
\vspace{-0.2in}
\end{table}

\subsection{System Overhead}
\label{subsec:overhead}
Tab.~\ref{tab:overhead} summarizes the per-sample inference overhead of \SystemName. The online overhead is dominated by the two foundation-model calls: agentic council reasoning contributes the largest latency at 5.442\,s per sample, followed by VLM-based semantic observation at 1.800\,s, while local retrieval takes only 0.735\,ms. The VLM and LLM stages consume 1.240K and 1.743K tokens per sample, respectively, yielding a total estimated cost of \$0.838 per 1K samples. These results indicate that structured multi-role reasoning is the dominant source of online overhead.

Overall, the end-to-end online latency is 7.243\,s. This overhead is suitable for segment-level activity understanding and offline analysis, but remains a limitation for latency-sensitive applications. Importantly, the zero-gradient self-evolution procedure is performed only once offline and therefore does not contribute to deployment-time per-sample overhead.

\section{Related Work}
\label{sec:related}

\subsection{mmWave-Based HAR}
\label{subsec:related_mmwave_har}
mmWave-based HAR~\cite{zhang2023survey} has been widely studied due to its privacy-preserving nature and robustness to lighting conditions. 
Prior work formulates mmWave HAR as a supervised classification problem over radar-derived representations such as time-frequency signatures~\cite{kim2022radar,feng2025mmwave,zhao2018rf} and point clouds~\cite{hao2025mm,qian20203d,ding2024milliflow}. 
Despite strong performance in fixed settings, prior studies have shown that learned representations are often sensitive to environmental factors and deployment conditions~\cite{jiang2018towards,chen2023cross}, making generalization under user, scene, and sensor variation a persistent challenge~\cite{liu2022mtranssee,zhou2025dgsense}.

To improve robustness under domain shift, subsequent efforts explored transfer learning, cross-modal knowledge transfer, and data synthesis. 
For example, mTransSee~\cite{liu2022mtranssee} improves environment-independent mmWave gesture recognition via transfer learning, RF-CM~\cite{wang2023rf} transfers cross-modal RF knowledge to support few-shot HAR, RFusion~\cite{feng2026rfusion} reduces data dependence through multimodal RF fusion, and RF Genesis~\cite{chen2023rfgenesis} improves generalization through simulation-based and generative data synthesis. Although these methods reduce annotation cost or improve cross-domain robustness, they still largely follow the paradigm of training or adapting a stronger recognition model for the target setting.

More recently, foundation-model-assisted approaches have emerged for mmWave HAR. mmCLIP~\cite{cao2024mmclip} aligns mmWave signals with text semantics for zero-shot HAR, RadarLLM~\cite{lai2025radarllm} extends radar sensing toward language-grounded motion understanding, and mmExpert~\cite{yan2025mmexpert} explores mmWave-text generation and interpretation with LLMs. However, these methods still rely mainly on learned semantic embeddings or open-ended language generation, with limited explicit physical grounding and weak decision auditability. In contrast, \SystemName targets deployment-time training-free mmWave HAR by grounding recognition in reusable radar knowledge, explicit physical evidence, and constrained multi-role reasoning.

\vspace{-0.15in}
\subsection{LLM-Augmented Sensing and Retrieval}
\label{subsec:related_llm_retrieval}

Recent work has increasingly introduced LLMs into sensor-based recognition by textualizing sensor signals or using language as an intermediate modality for transfer. 
Penetrative AI~\cite{xu2024penetrativeai} first showed that LLMs can reason over textualized physical-world sensor data. This paradigm was later extended to HAR through language-based cross-modality transfer, as explored in IMUGPT~\cite{leng2024imugpt} and IMUGPT 2.0~\cite{leng2024imugpt2}. 
LanHAR~\cite{yan2025lanhar}, SensorLLM~\cite{li2025sensorllm}, SensorLM~\cite{zhang2025sensorlm}, and LLaSA~\cite{asif2025llasa} further demonstrate the promise of LLMs for semantic abstraction, transfer, and open-vocabulary sensor reasoning. 
However, most of these methods still center on language alignment, synthetic supervision, or direct LLM-based recognition, rather than explicit physical validation.

Meanwhile, retrieval-augmented inference has begun to enter sensing and time-series analysis. Retrieval-augmented generation (RAG)~\cite{gao2023retrieval} is widely recognized as an effective mechanism for grounding LLM outputs in external evidence and mitigating hallucinations~\cite{li2025mitigating}. 
In sensing, RAG-HAR~\cite{sivaroopan2025raghar} augments HAR with retrieved reference examples and LLM-based classification, while ZARA~\cite{li2025zara} pushes this trend further toward retrieval- and agent-driven motion time-series reasoning. 
However, most of these methods are developed for IMU or more general sensor time-series data rather than RF sensing, and they primarily ground inference in textual knowledge, semantic exemplars, statistical descriptors, or learned embedding spaces, with limited explicit physical verification. 
In contrast, \SystemName targets deployment-time training-free mmWave HAR, performs retrieval in a physics-driven numerical feature space, and further enforces kinematic consistency through a dedicated Physicist role within structured council reasoning.

\vspace{-0.05in}
\section{Discussion}
\label{sec:discussion}

\textbf{Toward Evidence-Grounded mmWave HAR.}
Existing supervised methods have substantially advanced mmWave HAR and remain effective when labeled data and retraining are available~\cite{zhang2023survey}. Rather than replacing them, \SystemName offers a complementary deployment-time paradigm that resolves activities through retrieved precedents, semantic observations, and physical validation, producing an auditable reasoning trail rather than only a black-box prediction.

\noindent\textbf{Deployment Trade-offs in Practice.}
The main practical trade-off of \SystemName is that its online latency and cost are dominated by foundation-model inference. As a result, it is better suited to segment-level activity understanding than to highly time-sensitive control. At the same time, the framework can directly benefit from continued advances in VLM and LLM capability, latency, and cost efficiency~\cite{jiang2025thunderserve,shekhar2024towards}.

\noindent\textbf{Scope and Boundary Conditions.}
\SystemName is deployment-time training-free, but still requires offline knowledge-base construction. It currently focuses on single-user, coarse-grained full-body activities and does not cover multi-person scenes or fine-grained motion primitives. 
The pipeline is closely tied to mmWave sensing, whose motion signatures are often clearer than those of lower-resolution RF modalities such as Wi-Fi~\cite{zhang2021widar3,wang2014eyes,gao2021towards}. Extending it to other RF settings may require redesigning the physical and semantic evidence.

\noindent\textbf{Future Directions.}
Promising directions include more scalable knowledge-base construction~\cite{ahuja2021vid2doppler,rahman2024mmvr,yang2023xgait}, extension to multi-person scenarios~\cite{zeng2022multi}, finer-grained motion modeling~\cite{li2022towards}, and more efficient inference via selective model invocation or lightweight local models~\cite{chen2023frugalgpt,dekoninck2024unified}. 

\vspace{-0.05in}

\section{Conclusion}
\label{sec:conclusion}

We presented \SystemName, a deployment-time training-free framework for mmWave human activity recognition that shifts the paradigm from deployment-specific model optimization to evidence-grounded inference over reusable radar knowledge.
By grounding decisions in physically comparable precedents, explicit kinematic evidence, and structured multi-role reasoning, \SystemName enables robust recognition without deployment-time adaptation or manual radar annotation across diverse scenarios.
More broadly, our results suggest that robust and auditable mmWave HAR may be achieved not only by training better recognizers, but also by reusing historical knowledge, explicit physical evidence, and structured reasoning at inference time.

\newpage

\bibliographystyle{ACM-Reference-Format}
\bibliography{ref}

@inproceedings{zhao2020m,
  title={M-Cube: A millimeter-wave massive MIMO software radio},
  author={Zhao, Renjie and Woodford, Timothy and Wei, Teng and Qian, Kun and Zhang, Xinyu},
  booktitle={Proceedings of the Annual International Conference on Mobile Computing and Networking (MobiCom)},
  pages={1--14},
  year={2020}
}

@inproceedings{zheng2024enhancing,
  title={Enhancing mmWave radar sensing using a phased-MIMO architecture},
  author={Zheng, Kai and Zhao, Wuqiong and Woodford, Timothy and Zhao, Renjie and Zhang, Xinyu and Hua, Yingbo},
  booktitle={Proceedings of the Annual International Conference on Mobile Systems, Applications and Services (MobiSys)},
  pages={56--69},
  year={2024}
}

@inproceedings{wang2014eyes,
  title={E-Eyes: Device-free location-oriented activity identification using fine-grained WiFi signatures},
  author={Wang, Yan and Liu, Jian and Chen, Yingying and Gruteser, Marco and Yang, Jie and Liu, Hongbo},
  booktitle={Proceedings of the Annual International Conference on Mobile Computing and Networking (MobiCom)},
  pages={617--628},
  year={2014}
}

@article{li2025mitigating,
  title={Mitigating hallucination in large language models (LLMs): An application-oriented survey on RAG, reasoning, and agentic systems},
  author={Li, Yihan and Fu, Xiyuan and Verma, Ghanshyam and Buitelaar, Paul and Liu, Mingming},
  journal={arXiv preprint arXiv:2510.24476},
  year={2025}
}

@article{chen2023cross,
  title={Cross-domain WiFi sensing with channel state information: A survey},
  author={Chen, Chen and Zhou, Gang and Lin, Youfang},
  journal={ACM Computing Surveys},
  volume={55},
  number={11},
  pages={1--37},
  year={2023},
  publisher={ACM}
}

@inproceedings{chen2023rfgenesis,
  title     = {RF Genesis: Zero-shot generalization of mmWave sensing through simulation-based data synthesis and generative diffusion models},
  author    = {Chen, Xingyu and Zhang, Xinyu},
  booktitle = {Proceedings of the ACM Conference on Embedded Networked Sensor Systems (SenSys)},
  year      = {2023},
  pages     = {1--14},
  publisher = {ACM}
}

@inproceedings{xu2024penetrativeai,
author = {Xu, Huatao and Han, Liying and Yang, Qirui and Li, Mo and Srivastava, Mani},
title = {Penetrative AI: Making LLMs comprehend the physical world },
year = {2024},
publisher = {ACM},
booktitle = {Proceedings of the International Workshop on Mobile Computing Systems and Applications (HotMobile)},
pages = {1–7},
numpages = {7}
}

@article{zhou2025dgsense,
  title   = {{DGSense}: A Domain Generalization Framework for Wireless Sensing},
  author  = {Zhou, Rui and Cheng, Yu and Li, Songlin and Zhang, Hongwang and Liu, Chenxu},
  journal = {arXiv preprint arXiv:2502.08155},
  year    = {2025}
}

@article{gu2021survey,
  title={A survey on deep learning for human activity recognition},
  author={Gu, Fuqiang and Chung, Mu-Huan and Chignell, Mark and Valaee, Shahrokh and Zhou, Baoding and Liu, Xue},
  journal={ACM Computing Surveys},
  volume={54},
  number={8},
  pages={1--34},
  year={2021},
  publisher={ACM New York, NY}
}

@article{beddiar2020vision,
  title={Vision-based human activity recognition: A survey},
  author={Beddiar, Djamila Romaissa and Nini, Brahim and Sabokrou, Mohammad and Hadid, Abdenour},
  journal={Multimedia Tools and Applications},
  volume={79},
  number={41},
  pages={30509--30555},
  year={2020},
  publisher={Springer}
}

@article{zhang2023survey,
  title={A survey of mmWave-based human sensing: Technology, platforms and applications},
  author={Zhang, Jia and Xi, Rui and He, Yuan and Sun, Yimiao and Guo, Xiuzhen and Wang, Weiguo and Na, Xin and Liu, Yunhao and Shi, Zhenguo and Gu, Tao},
  journal={IEEE Communications Surveys \& Tutorials},
  volume={25},
  number={4},
  pages={2052--2087},
  year={2023},
  publisher={IEEE}
}

@article{hao2025mm,
  title={mm-ARnet: Exploring millimeter wave radar point clouds for human action recognition},
  author={Hao, Zhanjun and Xiao, Jiaxing and Wang, Yuejiao and Wang, Guowei and Li, Fenfang},
  journal={IEEE Transactions on Mobile Computing},
  year={2025},
  publisher={IEEE}
}

@inproceedings{ding2024milliflow,
  title={milliFlow: Scene flow estimation on mmwave radar point cloud for human motion sensing},
  author={Ding, Fangqiang and Luo, Zhen and Zhao, Peijun and Lu, Chris Xiaoxuan},
  booktitle={European Conference on Computer Vision (ECCV)},
  pages={202--221},
  year={2024},
  organization={Springer}
}

@article{kim2022radar,
  title={Radar-based human activity recognition combining range--time--Doppler maps and range-distributed-convolutional neural networks},
  author={Kim, Won-Yeol and Seo, Dong-Hoan},
  journal={IEEE Transactions on Geoscience and Remote Sensing},
  volume={60},
  pages={1--11},
  year={2022},
  publisher={IEEE}
}

@article{kong2024survey,
  title={A survey of mmWave radar-based sensing in autonomous vehicles, smart homes and industry},
  author={Kong, Hao and Huang, Cheng and Yu, Jiadi and Shen, Xuemin},
  journal={IEEE Communications Surveys \& Tutorials},
  volume={27},
  number={1},
  pages={463--508},
  year={2024},
  publisher={IEEE}
}

@article{feng2025mmwave,
  title={mmWave radar-based unsupervised gesture recognition via image-aligned heterogeneous domain transfer},
  author={Feng, Qihua and Cheng, Kunpeng and Duan, Chunhui},
  journal={IEEE Transactions on Mobile Computing},
  year={2025},
  publisher={IEEE}
}

@article{hu2025human,
  title={Human activity recognition trained on simulated millimeter-wave radar data with domain adaptation},
  author={Hu, Yuxuan and Yang, Xianghan and Xia, Zhaoyang and Xu, Feng},
  journal={IEEE Transactions on Instrumentation and Measurement},
  year={2025},
  publisher={IEEE}
}

@article{alam2022vision,
  title={Vision-based human fall detection systems using deep learning: A review},
  author={Alam, Ekram and Sufian, Abu and Dutta, Paramartha and Leo, Marco},
  journal={Computers in Biology and Medicine},
  volume={146},
  pages={105626},
  year={2022},
  publisher={Elsevier}
}

@article{liao2020review,
  title={A review of computational approaches for evaluation of rehabilitation exercises},
  author={Liao, Yalin and Vakanski, Aleksandar and Xian, Min and Paul, David and Baker, Russell},
  journal={Computers in Biology and Medicine},
  volume={119},
  pages={103687},
  year={2020},
  publisher={Elsevier}
}

@article{debes2016monitoring,
  title={Monitoring activities of daily living in smart homes: Understanding human behavior},
  author={Debes, Christian and Merentitis, Andreas and Sukhanov, Sergey and Niessen, Maria and Frangiadakis, Nikolaos and Bauer, Alexander},
  journal={IEEE Signal Processing Magazine},
  volume={33},
  number={2},
  pages={81--94},
  year={2016},
  publisher={IEEE}
}

@article{st1989analysis,
  title={Analysis of variance (ANOVA)},
  author={St, Lars and Wold, Svante and others},
  journal={Chemometrics and Intelligent Laboratory Systems},
  volume={6},
  number={4},
  pages={259--272},
  year={1989},
  publisher={Elsevier}
}

@inproceedings{zhang2022synthesized,
  title={Synthesized millimeter-waves for human motion sensing},
  author={Zhang, Xiaotong and Li, Zhenjiang and Zhang, Jin},
  booktitle={Proceedings of the ACM Conference on Embedded Networked Sensor Systems (SenSys)},
  pages={377--390},
  year={2022}
}

@article{feng2026rfusion,
  title={RFusion: Dynamic multimodal RF fusion for few-shot human activity recognition},
  author={Feng, Chao and Chen, Jiashen and Liang, Shuo and Peng, Xiaopeng and Yang, Baizhou and Wang, Xuan and Huang, Zexuan and Meng, Xianjia and Chen, Xiaojiang},
  journal={IEEE Transactions on Mobile Computing},
  year={2026},
  publisher={IEEE}
}

@article{wang2023rf,
  title={RF-CM: Cross-modal framework for RF-enabled few-shot human activity recognition},
  author={Wang, Xuan and Liu, Tong and Feng, Chao and Fang, Dingyi and Chen, Xiaojiang},
  journal={Proceedings of the ACM on Interactive, Mobile, Wearable and Ubiquitous Technologies},
  volume={7},
  number={1},
  pages={1--28},
  year={2023},
  publisher={ACM New York, NY, USA}
}

@inproceedings{yan2025mmexpert,
  title={mmExpert: Integrating Large Language Models for Comprehensive mmWave Data Synthesis and Understanding},
  author={Yan, Yifan and Yang, Shuai and Guo, Xiuzhen and Wang, Xiangguang and Chow, Wei and Shu, Yuanchao and He, Shibo},
  booktitle={Proceedings of the Twenty-sixth International Symposium on Theory, Algorithmic Foundations, and Protocol Design for Mobile Networks and Mobile Computing},
  pages={1--10},
  year={2025}
}

@article{lai2025radarllm,
  title={RadarLLM: Empowering large language models to understand human motion from millimeter-wave point cloud sequence},
  author={Lai, Zengyuan and Yang, Jiarui and Xia, Songpengcheng and Lin, Lizhou and Sun, Lan and Wang, Renwen and Liu, Jianran and Wu, Qi and Pei, Ling},
  journal={arXiv preprint arXiv:2504.09862},
  year={2025}
}

@article{liu2022mtranssee,
  title={mTransSee: Enabling environment-independent mmWave sensing based gesture recognition via transfer learning},
  author={Liu, Haipeng and Cui, Kening and Hu, Kaiyuan and Wang, Yuheng and Zhou, Anfu and Liu, Liang and Ma, Huadong},
  journal={Proceedings of the ACM on Interactive, Mobile, Wearable and Ubiquitous Technologies},
  volume={6},
  number={1},
  pages={1--28},
  year={2022},
  publisher={ACM New York, NY, USA}
}

@inproceedings{cao2024mmclip,
  title={mmCLIP: Boosting mmWave-based zero-shot HAR via signal-text alignment},
  author={Cao, Qiming and Xue, Hongfei and Liu, Tianci and Wang, Xingchen and Wang, Haoyu and Zhang, Xincheng and Su, Lu},
  booktitle={Proceedings of the ACM Conference on Embedded Networked Sensor Systems (SenSys)},
  year={2024}
}

@inproceedings{jiang2018towards,
  title={Towards environment independent device free human activity recognition},
  author={Jiang, Wenjun and Miao, Chenglin and Ma, Fenglong and Yao, Shuochao and Wang, Yaqing and Yuan, Ye and Xue, Hongfei and Song, Chen and Ma, Xin and Koutsonikolas, Dimitrios and others},
  booktitle={Proceedings of the Annual International Conference on Mobile Computing and Networking (MobiCom)},
  pages={289--304},
  year={2018}
}

@article{leng2024imugpt,
  title={IMUGPT: Zero-shot IMU-based human activity understanding with LLMs},
  author={Leng, Zikang and others},
  year={2024},
  journal={arXiv:2401.01234}
}

@article{leng2024imugpt2,
  title={IMUGPT 2.0: Language-based cross modality transfer for sensor-based human activity recognition},
  author={Leng, Zikang and Bhatt, Amitrajit and Fetahaj, Hymalai and Gao, Alanson and Guzman-Nateras, Lourdes and Dong, Rui and others},
  journal={Proceedings of the ACM on Interactive, Mobile, Wearable and Ubiquitous Technologies},
  volume={8},
  number={3},
  pages={1--32},
  year={2024},
  publisher={ACM}
}

@article{yan2025lanhar,
  title     = {{LanHAR}: Large language model-guided semantic alignment for human activity recognition},
  author    = {Yan, Hua and Tan, Heng and Ding, Yi and Zhou, Pengfei and Namboodiri, Vinod and Yang, Yu},
  journal   = {Proceedings of the ACM on Interactive, Mobile, Wearable and Ubiquitous Technologies},
  volume    = {9},
  number    = {4},
  pages     = {1--25},
  year      = {2025},
  publisher = {ACM}
}

@inproceedings{asif2025llasa,
  title={LLaSA: A sensor-aware LLM for natural language reasoning of human activity from IMU data},
  author={Asif Imran Shouborno, Sheikh and Khan, Mohammad Nur Hossain and Biswas, Subrata and Islam, Bashima},
  booktitle={Companion of the ACM International Joint Conference on Pervasive and Ubiquitous Computing (UbiComp Companion)},
  pages={893--899},
  year={2025}
}

@article{sivaroopan2025raghar,
  title={RAG-HAR: Retrieval augmented generation-based human activity recognition},
  author={Sivaroopan, Nirhoshan and others},
  journal={arXiv preprint arXiv:2512.08984},
  year={2025}
}

@article{gao2023retrieval,
  title={Retrieval-augmented generation for large language models: A survey},
  author={Gao, Yunfan and Xiong, Yun and Gao, Xinyu and Jia, Kangxiang and Pan, Jinliu and Bi, Yuxi and Dai, Yixin and Sun, Jiawei and Wang, Haofen and Wang, Haofen},
  journal={arXiv preprint arXiv:2312.10997},
  volume={2},
  number={1},
  year={2023}
}

@inproceedings{li2025sensorllm,
  title={Sensorllm: Aligning large language models with motion sensors for human activity recognition},
  author={Li, Zechen and Deldari, Shohreh and Chen, Linyao and Xue, Hao and Salim, Flora D},
  booktitle={Proceedings of the Conference on Empirical Methods in Natural Language Processing (EMNLP)},
  pages={354--379},
  year={2025}
}

@inproceedings{zhang2025sensorlm,
  title     = {{SensorLM}: Learning the language of wearable sensors},
  author    = {Zhang, Yuwei and Ayush, Kumar and Qiao, Siyuan and Heydari, A. Ali and Narayanswamy, Girish and Xu, Maxwell A. and Metwally, Ahmed A. and Xu, Shawn and Garrison, Jake and Xu, Xuhai and Althoff, Tim and Liu, Yun and Kohli, Pushmeet and Zhan, Jiening and Malhotra, Mark and Patel, Shwetak and Mascolo, Cecilia and Liu, Xin and McDuff, Daniel and Yang, Yuzhe},
  booktitle = {Advances in Neural Information Processing Systems},
  year      = {2025}
}

@article{li2025zara,
  title   = {{ZARA}: Zero-shot motion time-series analysis via knowledge and retrieval driven {LLM} agents},
  author  = {Li, Zechen and Chen, Baiyu and Xue, Hao and Salim, Flora D.},
  journal = {arXiv preprint arXiv:2508.04038},
  year    = {2025}
}

@article{shekhar2024towards,
  title={Towards optimizing the costs of LLM usage},
  author={Shekhar, Shivanshu and Dubey, Tanishq and Mukherjee, Koyel and Saxena, Apoorv and Tyagi, Atharv and Kotla, Nishanth},
  journal={arXiv preprint arXiv:2402.01742},
  year={2024}
}

@inproceedings{jiang2025thunderserve,
  title={Thunderserve: High-performance and cost-efficient llm serving in cloud environments},
  author={Jiang, Youhe and Fu, Fangcheng and Yao, Xiaozhe and Wang, Taiyi and Cui, Bin and Klimovic, Ana and Yoneki, Eiko},
  journal={Proceedings of Machine Learning and Systems (MLSys)},
  year={2025}
}

@inproceedings{zeng2022multi,
  title={Multi-HAR: Human activity recognition in multi-person scenes based on mmWave sensing},
  author={Zeng, Xianlin and Shi, Yiming and Zhou, Anfu},
  booktitle={Proceedings of the IEEE International Conference on Computer and Communications (ICCC)},
  pages={1789--1793},
  year={2022},
  organization={IEEE}
}

@article{li2022towards,
  title={Towards domain-independent and real-time gesture recognition using mmWave signal},
  author={Li, Yadong and Zhang, Dongheng and Chen, Jinbo and Wan, Jinwei and Zhang, Dong and Hu, Yang and Sun, Qibin and Chen, Yan},
  journal={IEEE Transactions on Mobile Computing},
  volume={22},
  number={12},
  pages={7355--7369},
  year={2022},
  publisher={IEEE}
}

@inproceedings{ahuja2021vid2doppler,
  title={Vid2Doppler: Synthesizing doppler radar data from videos for training privacy-preserving activity recognition},
  author={Ahuja, Karan and Jiang, Yue and Goel, Mayank and Harrison, Chris},
  booktitle={Proceedings of the CHI Conference on Human Factors in Computing Systems (CHI)},
  pages={1--10},
  year={2021}
}

@inproceedings{rahman2024mmvr,
  title={MMVR: Millimeter-wave multi-view radar dataset and benchmark for indoor perception},
  author={Rahman, M Mahbubur and Yataka, Ryoma and Kato, Sorachi and Wang, Pu and Li, Peizhao and Cardace, Adriano and Boufounos, Petros},
  booktitle={European Conference on Computer Vision (ECCV)},
  pages={306--322},
  year={2024},
  organization={Springer}
}

@inproceedings{yang2023xgait,
  title={XGait: Cross-modal translation via deep generative sensing for RF-based gait recognition},
  author={Yang, Huanqi and Han, Mingda and Jia, Mingda and Sun, Zehua and Hu, Pengfei and Zhang, Yu and Gu, Tao and Xu, Weitao},
  booktitle={Proceedings of the ACM Conference on Embedded Networked Sensor Systems (SenSys)},
  pages={43--55},
  year={2023}
}

@article{chen2023frugalgpt,
  title={Frugalgpt: How to use large language models while reducing cost and improving performance},
  author={Chen, Lingjiao and Zaharia, Matei and Zou, James},
  journal={arXiv preprint arXiv:2305.05176},
  year={2023}
}

@article{dekoninck2024unified,
  title={A unified approach to routing and cascading for llms},
  author={Dekoninck, Jasper and Baader, Maximilian and Vechev, Martin},
  journal={arXiv preprint arXiv:2410.10347},
  year={2024}
}

@misc{TI_AWR1843,
  title        = {AWR1843 Single‑Chip 76 {GHz} to 81 {GHz} Automotive Radar Sensor},
  howpublished = {\url{https://www.ti.com/product/AWR1843}},
  year         = {2025},
  author       = {{Texas Instruments}},
}

@misc{TI_DCA1000,
  title        = {DCA1000 Evaluation Module for Real-time Data Capture and Streaming},
  howpublished = {\url{https://www.ti.com/tool/DCA1000EVM}},
  year         = {2025},
  author       = {{Texas Instruments}},
}

@misc{Hikv,
  title        = {DS-E12 2MP USB Camera},
  howpublished = {\url{https://dealer-static.hikvision.com/upload/file/doc/DOC000065542-DS-E12.pdf}},
  year         = {2025},
  author       = {{Hangzhou Hikvision}},
}

@misc{caaresys,
title        = {Caaresys},
author       = {{Wikipedia}},
year         = {2026},
howpublished = {\url{https://en.wikipedia.org/wiki/Caaresys}}
}

@misc{aqara, 
title = {Aqara's cord-free presence sensor runs for up to three years on battery power}, 
author = {Liszewski, Andrew}, 
year = {2025}, 
howpublished = {\url {https://www.theverge.com/news/819067/aqara-presence-multi-sensor-fp300-battery-power-zigbee-matter-thread}}
}

@article{qian20203d,
  title={3D point cloud generation with millimeter-wave radar},
  author={Qian, Kun and He, Zhaoyuan and Zhang, Xinyu},
  journal={Proceedings of the ACM on Interactive, Mobile, Wearable and Ubiquitous Technologies},
  volume={4},
  number={4},
  pages={1--23},
  year={2020},
  publisher={ACM New York, NY, USA}
}

@inproceedings{zhao2018rf,
  title={RF-based 3D skeletons},
  author={Zhao, Mingmin and Tian, Yonglong and Zhao, Hang and Alsheikh, Mohammad Abu and Li, Tianhong and Hristov, Rumen and Kabelac, Zachary and Katabi, Dina and Torralba, Antonio},
  booktitle={Proceedings of the Conference of the ACM special Interest Group on Data Communication (SIGCOMM)},
  pages={267--281},
  year={2018}
}

@article{gao2021towards,
  title={Towards position-independent sensing for gesture recognition with Wi-Fi},
  author={Gao, Ruiyang and Zhang, Mi and Zhang, Jie and Li, Yang and Yi, Enze and Wu, Dan and Wang, Leye and Zhang, Daqing},
  journal={Proceedings of the ACM on Interactive, Mobile, Wearable and Ubiquitous Technologies},
  volume={5},
  number={2},
  pages={1--28},
  year={2021},
  publisher={ACM New York, NY, USA}
}

@inproceedings{gong2025seradar,
  title={SeRadar: Embracing secondary reflections for human sensing with mmWave radar},
  author={Gong, Danei and Zheng, Naiyu and Xie, Binbin and Xiong, Jie and Wang, Shuai and Fang, Yuguang and Yin, Zhimeng},
  booktitle={Proceedings of the Annual International Conference on Mobile Computing and Networking (MobiCom)},
  pages={231--246},
  year={2025}
}

@article{zhang2021widar3,
  title={Widar3.0: Zero-effort cross-domain gesture recognition with Wi-Fi},
  author={Zhang, Yi and Zheng, Yue and Qian, Kun and Zhang, Guidong and Liu, Yunhao and Wu, Chenshu and Yang, Zheng},
  journal={IEEE Transactions on Pattern Analysis and Machine Intelligence},
  volume={44},
  number={11},
  pages={8671--8688},
  year={2021},
  publisher={IEEE}
}

@inproceedings{ding2020rf,
  title={RF-Net: A unified meta-learning framework for RF-enabled one-shot human activity recognition},
  author={Ding, Shuya and Chen, Zhe and Zheng, Tianyue and Luo, Jun},
  booktitle={Proceedings of the Conference on Embedded Networked Sensor Systems (SenSys)},
  pages={517--530},
  year={2020}
}

\end{document}